\documentclass[a4paper,11pt]{article}
\pdfoutput=1 

\usepackage{jheppub} 

\usepackage[T1]{fontenc} 
\usepackage{float}
\usepackage{comment}
\usepackage{color}
\usepackage{dcolumn}
\usepackage{paralist}
\usepackage{amsmath}
\usepackage{graphicx,epsfig}
\usepackage{caption}
\usepackage{pifont}
\usepackage{subcaption}
\usepackage{color,amsmath,amssymb,amsthm, ragged2e}
\usepackage{cancel}

\def\mEt{\mbox{${\hbox{$E$\kern-0.6em\lower-.1ex\hbox{/}}}_T$}\, } 

\def\snu{{\tilde\nu}}
\def\snuL{{\tilde\nu_L}}
\def\sN{{\tilde{N}}}
\def\sNR{{\tilde{N}_R}}

\newcommand{\newc}{\newcommand}

\newc{\mstop}{m_{\tilde{t}}}

\newc{\mtop}{m_t}
\newc{\mbot}{m_b}
\newc{\mz}{m_Z}
\newc{\mw}{M_W}

\newc{\sgn}{\mbox{sgn}}
\newc{\tbeta}{\tan\beta}

\newc{\Mlsp}{M_{\rm LSP}}

\def\beq{\begin{equation}}
\def\eeq{\end{equation}}
\def\bea{\begin{eqnarray}}
\def\eea{\end{eqnarray}}

\title{\boldmath  PeV scale Supersymmetry breaking and the IceCube neutrino flux}

 \author[a,1]{Mansi Dhuria,\note{Corresponding author.}}
 \author[a]{Vikram Rentala}

 \affiliation[a]{Department of Physics, Indian Institute of Technology Bombay, \\ Mumbai 400 076, India}

 \emailAdd{mansidh@phy.iitb.ac.in}
 \emailAdd{rentala@phy.iitb.ac.in}

\abstract{The observation of very high energy neutrino events at IceCube has grasped a lot of attention in the fields of both astrophysics and particle physics. It has been speculated that these high energy neutrinos might originate either from purely conventional astrophysical sources or from the late decay of a super heavy (PeV scale) dark matter (DM) particle. In order for decaying DM to be a dominant source of the IceCube high-energy neutrinos, it would require an unusually suppressed value of the coupling of DM to neutrinos. We attempt to explain this small coupling in the context of an $R$-parity conserving minimal supergravity model which has right-handed neutrino superfields. With the main assumptions of super-partner masses at the PeV scale and also a reheating temperature not much larger than the PeV scale, we find in our model several natural order-of-magnitude ``miracles'', (i) the gravitino is produced via freeze-in as a DM candidate with the correct relic density (ii) the right-handed (RH) sneutrino makes up only a tiny fraction ($10^{-6})$, of the present day energy density of the universe, yet its decay lifetime to the gravitino and neutrinos is such that it naturally predicts the right order-of-magnitude for the IceCube neutrino flux. The long lifetime of the RH sneutrino is explained by the existence of a global $R$-symmetry which is only broken due to supersymmetry breaking effects. Our model also predicts a flux of 100~TeV gamma rays from the decaying RH sneutrino which are within the current observational constraints.}

\begin{document}

\maketitle
\flushbottom


\section{Introduction}

For the past few decades, supersymmetry (SUSY) has been a dominant paradigm to address almost all the shortcomings of the Standard Model (SM) \cite{Dimopoulos:1981zb}. However, so far direct searches at the Large Hadron Collider (LHC)~\cite{ATLASSUSY:2017,CMSSUSY:2017} for weak scale supersymmetry have produced only null results, which has forced us to reconsider whether naturalness is the prime motivation for predicting the SUSY breaking scale. Indirect searches from flavor physics observables for generic SUSY breaking scenarios with large flavor mixings have long constrained the SUSY breaking scale to be greater than 100 TeV~\cite{Raby:2017ucc}. These strong constraint have forced theorists to work in scenarios with minimal flavor violation that are arguably less generic, such as gauge mediated SUSY breaking (see for example~\cite{Kolda:1997wt, Giudice:1998bp, Meade:2008wd} and references therein).

However, SUSY is still a theoretically attractive and predictive paradigm. If one takes the message from LHC and flavor physics searches seriously, then the SUSY breaking masses should lie beyond the 100 TeV scale. Even if SUSY does not solve the hierarchy problem, it has many other theoretically attractive features, such as the prediction of improved gauge coupling unification compared to the SM~\cite{Wells:2003tf, ArkaniHamed:2004fb, Giudice:2004tc} among other features.

One recent experimental observation which motivates the study of high scale SUSY breaking is the observation of PeV-scale neutrino events at the IceCube observatory~  \cite{Aartsen:2013jdh,Aartsen:2014gkd,Aartsen:2015knd,Aartsen:2015zva,kopper:2017}. The IceCube detector at the South Pole is designed to search for high energy cosmic-ray neutrinos. The IceCube collaboration has reported 82 ultra-high energy neutrino events corresponding to deposited energies in the range from 60 TeV to 10 PeV in their six year data sets~\cite{kopper:2017}. Since IceCube disfavors a purely atmospheric neutrino source as an explanation for these events~\cite{Aartsen:2013jdh}, the flux of these high energy neutrinos has been posited to originate from the decay of a heavy dark matter (DM) particle~
\cite{Esmaili:2012us,Esmaili:2013gha,Feldstein:2013kka,Esmaili:2014rma,Rott:2014kfa,Roland:2015yoa,Ko:2015nma,Ema:2016zzu,Fiorentin:2016avj,Dev:2016qbd,Chianese:2016smc,Boucenna:2015tra,Borah:2017xgm,Hiroshima:2017hmy,Bhattacharya:2017jaw,Chakravarty:2017hcy}
\footnote{The idea that decay of superheavy DM may lead to high energy cosmic ray signatures predates the IceCube data and has been studied in the literature in both non-supersymmetric \cite{Berezinsky:1997hy,Kuzmin:1997jua} and supersymmetric models \cite{Berezinsky:2008bg}.} or from astrophysical sources such as extragalactic Supernova Remnants (SNRs)~\cite{Murase:2013rfa,Tamborra:2014xia,Chakraborty:2015sta}, Hypernova Remnants~(HNRs)~\cite{Murase:2013rfa,Tamborra:2014xia,Chakraborty:2015sta}, Active Galactic Nuclei (AGNs)~\cite{ Murase:2014foa,Dermer:2014vaa,Kimura:2014jba,Kalashev:2014vya}, and gamma-ray bursts (GRBs)~\cite{Waxman:1997ti,Murase:2013ffa}.  We focus on a decaying DM interpretation of the high energy (PeV scale) IceCube neutrinos in this work.

If DM decay is the correct explanation for the high energy neutrino events, it forces us to consider a PeV scale decaying DM candidate in the theory. However, a PeV scale DM candidate cannot be produced via thermal freeze-out, as the cross-section would be limited by the unitarity bound~\cite{Griest:1989wd}. Thus a PeV-scale neutralino would not be an ideal candidate to explain the observed abundance of DM and simultaneously predict the IceCube neutrino flux. This observation compels us to consider other potential DM candidates e.g. gravitinos, axinos etc., which could be produced using a non-thermal mechanism to obtain the correct DM relic abundance. In addition to this, if the decay of DM is considered to be the dominant source of high-energy neutrinos, one requires an extremely high lifetime of the DM (dominantly decaying into one or more neutrinos) in order to predict a neutrino flux that is in good agreement with IceCube data. The requirement of the high lifetime $\sim\mathcal{O}(10^{28}-10^{30})$~seconds of DM translates into an unusually suppressed value of the DM-neutrino interaction strength which is very challenging to explain theoretically. Refs.~\cite{Ko:2015nma,Ema:2016zzu,Fiorentin:2016avj,Chianese:2016smc,Borah:2017xgm,Feldstein:2013kka,Dev:2016qbd} attempted to explain the observed neutrino flux by allowing for a fine-tuned value of the DM-neutrino interaction coupling in their models. There have however been a few attempts which try to explain the origin of a small value of the DM-neutrino coupling. In refs.~\cite{Feldstein:2013kka,Chakravarty:2017hcy}, the required coupling is obtained by considering tiny $R$-parity violating interactions which are ``technically natural''. One can also consider explicit continuous global symmetry breaking parameters that can be chosen to be small (e.g. ref.~\cite{Dev:2016qbd} introduces a small chiral symmetry breaking mass for an electron partner and the right chiral electron field). Ref.~\cite{Boucenna:2015tra,Roland:2015yoa} generate the small coupling through higher dimensional operators which are suppressed by some heavy mass scale, such as the Grand Unified Theory (GUT) scale or Planck scale.

In this work, we propose a simple ${\cal N}=1$ supergravity model which includes the particles of the Minimal Supersymmetric Standard Model (MSSM)~\cite{Martin:1997ns} and an additional right-handed (RH) neutrino superfield. Our main assumption is that the superpartner masses are at the PeV scale. The gravitino is assumed to be the lightest supersymmetric particle (LSP) and the RH sneutrino\footnote{As a reminder to the reader, the scalar component of the RH neutrino supermultiplet is the RH sneutrino.} is assumed to be the next-to-lightest superparticle (NLSP). Our model assumes the existence of a global $U(1)_R$-symmetry with specific $R$-charge assignments. A well known motivation for imposing such a global $R$-symmetry is that the breaking of this symmetry due to SUSY breaking effects can explain the small value of the $\mu$ parameter, hence addressing the infamous ``$\mu$-problem'' via the Giudice-Masiero mechanism~\cite{Giudice:1988yz}. We will argue that in the presence of $R$-parity and the additional $R$-symmetry, the RH sneutrino becomes a quasi-stable candidate with decays forbidden by the approximately preserved $R$-symmetry. However, after the breaking of $R$-symmetry at the PeV scale, the non-renormalizable operators involving the sneutrino allow for it to decay into a gravitino and other SM particles with an extremely long lifetime. Since the dominant decay mode of the sneutrino happens to produce a neutrino as one of the final decay products, it is an attractive candidate to explain the PeV scale neutrino events.

We will see several ``miracles'' that occur naturally under this set-up, (i) the gravitino is predicted to make up most of the dark matter and naturally has the correct order-of-magnitude relic abundance from freeze-in production, (ii) we will see that the sneutrino is predicted to contribute to less than $10^{-6}$ of the present day energy density of the universe, however the value of the decay lifetime will naturally result in a flux spectrum normalization that can explain the PeV scale neutrino events observed at IceCube. Furthermore, we will show that the same set-up can solve the $\mu$-problem and can also explain the small value of the neutrino masses.

This paper is organized as follows: In \textsection \ref{sect:outline}, we first discuss the details of our ${\cal N}=1$ supergravity model. We write down the leading $R$-symmetry conserving non-renormalizable operators appearing in the superpotential and K\"{a}hler potential for this model. In \textsection\ref{sneutrinomassmatrix}, we present the spectrum of superpartners along with some benchmark values and we work out the physical mass eigenstates of the sneutrino by diagonalizing the mass matrix involving off-diagonal mixing terms coming from soft SUSY breaking trilinear couplings. In \textsection\ref{Higgsmass}, we briefly discuss the conditions under which one can obtain a SM-like Higgs with PeV-scale SUSY. In \textsection\ref{neutrino}, we calculate the mass of the SM-like left-handed neutrino in this model, which turns out to be purely Majorana in nature. In \textsection\ref{sect:lifetime}, we calculate the decay width of the right-handed sneutrino and show that is has a lifetime longer than the age of the universe, indicating that it can act as a viable decaying DM candidate. In \textsection\ref{relic}, we calculate the relic abundance of both the gravitino and the sneutrino produced non-thermally from the scattering and decay of heavier supersymmetric particles. In \textsection\ref{icecube}, we compute the expected neutrino flux from the decay of sneutrino DM along with a simple unbroken power-law astrophysical source, and we show that the combination explains the IceCube data well. We also show that our results are consistent with current constraints from high-energy gamma ray observations. In \textsection\ref{icecube}, we summarize our results. In appendix \textsection \ref{appendix1}, we review the mechanism that leads to both $R$-symmetry and SUSY breaking in our model.

\section{Outline of the model}
\label{sect:outline}
In this section, we start by presenting the details of our supergravity model of PeV scale SUSY breaking. As mentioned in the introduction, the model assumes an additional global $U(1)_R$ symmetry. We will first discuss why such a $U(1)_R$ symmetry has been considered desirable in previous works.

Usually it is assumed that SUSY is broken in the hidden sector at a very high scale and SUSY breaking effects are transmitted to the visible sector via gravitational interactions. Therefore the coefficient of the SUSY breaking operators, such as the masses of superpartners, are determined by non-renormalizable operators governing the interactions between the visible and hidden sectors. Infamously, there is an ambiguous $\mu$ parameter present in the superpotential which does not depend on the dynamics of SUSY breaking. Its natural value can be either zero (if protected by some symmetry) or on the order of the Planck scale. However, the value of the $\mu$ parameter needs to be tuned to the order of the SUSY breaking scale (or smaller) in order for electroweak symmetry breaking (EWSB) to occur successfully. This is the well known ``$\mu$ problem'' in SUSY theories. A solution suggested by Giuidice and Maisero~\cite{Giudice:1988yz} in the context of ${\cal N}=1$ supergravity theories imposes an additional global $U(1)_R$ symmetry. Their idea was based on the simple observation that if the Higgs superfields are vector-like under the $U(1)_R$ symmetry, then the term $\mu H_u H_d$ would be forbidden in the superpotential, and EWSB can successfully occur. However, in order to generate higgsino masses, they also included Planck suppressed terms like $({X^\dagger}/{M_{\textrm{p}}}) H_u H_d $ in the K\"{a}hler potential, with $X$ being a hidden-sector superfield (SUSY breaking spurion) and $M_{\textrm{p}}$ being the Planck mass scale.

Subsequently, the Giuidice-Maisero mechanism was extended to involve a SM singlet RH neutrino in refs.~\cite{ArkaniHamed:2000bq, MarchRussell:2004uf,ArkaniHamed:2000kj} with the motivation of realizing a TeV scale see-saw mechanism of neutrino mass generation. This scenario predicted a TeV-scale Majorana neutrino which could be observed at the LHC. In these models, the charges of the right-handed neutrino superfields $(N^c)$ under the global $U(1)_R$ symmetry prevented Majorana mass terms from showing up in the superpotential (similar to the way the Higgs bilinear term is forbidden), but allowed for the non-renormalizable operator $({X^\dagger}/{M_{\textrm{p}}}) \widehat N^c \widehat N^c $ in the K\"{a}hler potential, thus giving rise to a right-handed neutrino mass on the order of the SUSY breaking scale. Interestingly, this mechanism explains the small value of the observed neutrino masses via the see-saw mechanism. In the model discussed in ref.~{\cite{ArkaniHamed:2000bq}}, the neutrino mass is generated from non-renormalizable terms in the K\"{a}hler potential while the right-handed neutrino Yukawa couplings and corresponding soft SUSY breaking trilinear coupling parameters are generated from other non-renormalizable operators in the superpotential.

We draw motivation from these models above, to build a model of PeV scale SUSY breaking which possesses a $U(1)_R$ symmetry and has  right-handed neutrino superfields in addition to the superfields of the MSSM. In order to generate a quasi-stable, long-lived right-handed sneutrino, we assign different $R$-charges to the fields following which all the soft SUSY breaking operators involving the RH neutrino superfield, including Yukawa terms, are generated from higher dimensional operators in the K\"{a}hler potential. By imposing the $R$-charges $R(\widehat N^c) = R( \widehat H_u) = R( \widehat H_d) = 1/2,~R( \widehat X)= 1,~R( \widehat L)= R( \widehat Q) = 0, R( \widehat e^c) = R( \widehat u^c) = R( \widehat d^c) =  3/2$,  one does not get any renormalizable terms in either the K\"{a}hler or superpotential involving direct interactions between RH neutrinos and other MSSM superfields. The non-vanishing interaction terms thus appear from the following $R$-symmetry conserving non-renormalizable operators in the K\"{a}hler potential:
\bea
\label{eq:K}
{\cal K} &= &  \frac{  \widehat X^\dagger}{M_{\textrm{p}}}\widehat H_u \widehat H_d  +\frac{\widehat X^\dagger}{M_{\textrm{p}}} \widehat N^c  \widehat N^c \left(1 + \frac{ \widehat X^\dagger  \widehat X}{M^2_{\textrm{p}}}\right)  +  \frac{  \widehat X^\dagger }{M^2_{\textrm{p}}} \widehat N^c \widehat L \cdot\widehat H_u  \left (1 + \frac{ \widehat X^\dagger  \widehat X}{M^2_{\textrm{p}}} + ..\right)  \nonumber\\
 && + \frac{ \widehat X^\dagger  \widehat X}{M^2_{\textrm{p}} }\left(\widehat L^{\dagger}  \widehat L  +  \widehat N^{c \dagger}\widehat N^c \right)  +  \frac{1}{M^2_*} \left(\widehat L \cdot \widehat H^{\dagger}_d \right) (\widehat L\cdot \widehat H_u) + \textrm{ h.c.},
\eea
where $M_{\textrm{p}}$ is the Planck scale and $M_{*}$ is the scale obtained by integrating out other heavy states (assumed to be present in the theory).

In the `Planck-scale mediated' SUSY breaking scenario, if SUSY is broken in the hidden sector by an $F$-term vacuum expectation value (VEV) $\langle F_X \rangle$, soft mass terms in the visible sector are of the form
\beq
m_{\rm soft } \sim  \frac{F_X} {M_{\textrm{p}}}.
\eeq
The spontaneous breaking of local supersymmetry implies the existence of a massive spin-$3/2$ gravitino. Its mass is also given by
\beq
m_{3/2} = \frac{F_X} {M_{\textrm{p}}}.
\eeq
In order for the sparticle masses to be at the PeV-scale, one needs to assume $F_X \sim 10^{12}$~PeV$^2$. This gives rise to soft SUSY breaking mass terms and a gravitino mass at the PeV scale.

The masses of the right-handed neutrino as well as the higgsino appear from the non-renormalizable operators given by the first two terms of the  K\"{a}hler potential. Upon considering the $F$-term VEV, the RH neutrino mass and the higgsino mass parameter ($\mu$) turn out to be of the same order,
\beq
m_{N} \sim \mu \sim  \frac{F_X^*} {M_{\textrm{p}}}   \sim  ~{\rm PeV}.
\eeq

The Yukawa coupling for the right-handed neutrino appears through the non-renormalizable operator $\frac{ \widehat X^\dagger }{M^2_P} \widehat N^c  \widehat L \widehat H_u$ in the K\"{a}hler potential. By writing the superfield in terms of component fields, we find that the Yukawa coupling is given by
\begin{equation}
 y_N =  \alpha_N \frac{F_X^*} {M^2_{\textrm{p}}}  \sim 10^{-13},
\end{equation}
for an assumed coupling $\alpha_N \sim {\cal O}(0.1)$. Further we can see that the operator $({ \widehat X^\dagger }/{M^2_{\textrm{p}}}) \widehat N^c  \widehat L \widehat H_u$ in the K\"{a}hler potential does not produce any trilinear scalar coupling. However, the trilinear term can arise from the sub-leading non-renormalizable operator given by  $\frac{ (\widehat X \cdot  \widehat X^\dagger) \widehat X^\dagger }{M^4_p} \widehat N^c \widehat L \widehat H_u$.  For  $ \langle X \rangle \sim m_{3/2}$ (a justification for expecting this particular VEV for the scalar component of the SUSY breaking spurion field is given in appendix {\bf A}) and $F_{X} \sim 10^{12} $ PeV$^2$,  ${A}_{N}$ is given by
\begin{equation}
{ A}_{N} \sim  \beta_N \frac{\langle X \rangle }{M^2_{\textrm{p}}} \times \left(\frac{F^*_X  F_X} {M^2_{\textrm{p}}}\right) \sim 10^{-24} \  {\rm PeV},
\end{equation}
for an assumed coupling $\beta_N \sim {\cal O}(1)$. Similarly the lepton-number violating $B$-term coefficient $B_N$ (which is the coefficient of $NN$ in the Lagrangian) as well the Higgs mixing parameter $B_\mu$ (which is the coefficient of $H_u H_d$ in the Lagrangian), will be given by
\bea
B_N \sim B_\mu \sim  \frac{\langle X \rangle }{M_{\textrm{p}}} \times \left(\frac{F^*_X  F_X} {M^2_{\textrm{p}}}\right) \sim 10^{-12} \  {\rm PeV^2}.
\eea

We note that in addition to the soft SUSY breaking terms we have estimated, we also need a mechanism to generate gaugino masses and slepton/squark trilinear interaction terms ($A$-terms). We will not discuss in detail the mechanism for generation of these terms, but one way these could be generated is by assuming that in addition to hidden-sector field $X$ with $R$-charge $R_X =1$ as considered in this paper, there could be another hidden sector superfield $Y$ present in the theory with $R_Y=0$. The non-renormalizable interactions of this field (in both the K\"{a}hler potential and superpotential) would generate the gaugino masses and trilinear terms to be of the order of PeV. We note that this mechanism would \emph{not} however, generate a trilinear term involving the RH sneutrino ($A_N$) because the $R$-charge of the sneutrino superfield forbids the corresponding K\"{a}hler terms.

\paragraph{Effect of one-loop corrections:} It is evident that values of all the parameters we have estimated are at the hidden sector SUSY breaking scale ($\sqrt{F_X} \sim 10^6$ PeV). In order to obtain their physical values at the low energy (EW/PeV) scale, one has to consider renormalization group (RG) evolution of the mass parameters and couplings. Since all the mass parameters in our model (except $B_{\mu}$ and $B_{N}$) are around the PeV-scale, it is reasonable to assume that there will only be an ${\cal O}(1)$ change in their values after RG evolution down to the low energy scale. Similarly, we expect that all the couplings except the suppressed ones involving right-handed neutrinos ($y_N$ and $A_N$) will exhibit an ${\cal O}(1)$ change after their RG evolution down to the EW scale.

Next we check whether the suppressed parameters $y_N$, $A_N$, $B_{\mu}$, $B_{N}$ get enhanced at the one-loop level due to other dominant couplings and PeV-scale mass parameters present in the model. Using the generic expressions for one-loop RG equations as given in \cite{Martin:1997ns}, we have checked that the value of Yukawa coupling $y_N$ will remain mostly unchanged after its RG evolution down to the low energy scale. However, the radiatively corrected trilinear scalar coupling $A_N$ will receive a dominant correction proportional to the Yukawa coupling $y_N$ given by,

\beq
\delta {A_N}_{(\rm 1-loop)} \propto  \frac{y_N}{16 \pi^2} \left( g^2_a M_a  \right) \sim 10^{-15} \textrm{ PeV}.
\eeq
 The numerical estimate above holds if we take $y_N \sim 10^{-13}$ and the electroweakino mass $M_a \sim {\rm PeV}$, with $g_a$ being the corresponding gauge coupling. Similarly, by using the generic expressions for one-loop RG equations as given in \cite{Martin:1997ns}, it can be checked that the value of $B_N$ will not get any significant contribution even at one-loop. However, the radiatively corrected soft SUSY breaking mixing parameter $B_\mu$ receives a dominant contribution given by
\beq
\delta {B_\mu}_{(\rm 1-loop)} \propto \frac{\mu}{16 \pi^2} \left( g^2_a M_a  \right).
\eeq
The values of the higgsino mass parameter ($\mu$) and the electroweak gaugino masses ($M_a$) can be chosen in such a way that the one-loop corrected $B_{\mu}$ turns out to be on the order of the PeV scale.

Thus in summary, this model gives PeV scale mass to all supersymmetric partners. The imposition of $R$-symmetry gives rise to suppressed values of both $y_N$ and ${A}_N$, while the trilinear couplings related to all other supersymmetric partners will still remain on the order of the PeV scale.

\begin{table}
\begin{center}
\begin{tabular}{ |c|c|c| }
 \hline
Parameters & Dependence  & Numerical values \\
& on $\langle X \rangle $ and $ F_X $ &\\
 \hline
Scalar/squark mass ($m_i$) & $F_X/M_{\textrm{p}}$  & ${\cal O}$(1) PeV \\
 \hline
Higgsino mass parameter ($\mu$) & $F_X/M_{\textrm{p}}$ & ${\cal O}$(1)  PeV  \\
 \hline
 Higgs mixing parameter ($B_\mu$) & $\frac{\mu}{16 \pi^2} \left( g^2_a M_a  \right)$ & ${\cal O}$(1)  PeV  \\
 \hline
RH neutrino mass ($m_N$)&$F_X/M_{\textrm{p}}$ & ${\cal O}$(1) PeV \\
 \hline
Lepton number violating B-term ($B_N$)& ${\langle X \rangle F^*_X  F_X} /{M^3_p}$ & ${\cal O}(10^{-12})~{\rm PeV}^2$ \\
 \hline
Trilinear scalar couplings ($A_i$)  & $F_X/M_{\textrm{p}}$ &${\cal O}$(1)  PeV \\
 (except involving sneutrino) & & \\
 \hline
Sneutrino trilinear scalar coupling ($A_N$) & $\frac{y_N M_a}{16 \pi^2}  $ & ${\cal O}(10^{-15})$  PeV \\
 \hline
Neutrino Yukawa coupling ($y_N$) & ${F_X^*}/{M^2_{\textrm{p}}}$ &${\cal O}(10^{-13})$ \\
 \hline
 \end{tabular}
\caption{Table showing the order-of-magnitude of the SUSY breaking parameters and the RH neutrino/sneutrino couplings in our model. Note that all parameters must be evaluated at the SUSY breaking scale and must be renormalization group (RG) evolved to the low energy scale to find their physical values. While we expect only an $\mathcal{O}(1)$ change in most parameters, the only term that gets a large loop correction from RG evolution is the sneutrino trilinear scalar coupling $A_N$. The values shown in the table reflect the RG evolved values of all parameters evaluated at the appropriate physical scale.}
\label{table1}
\end{center}
\end{table}

Table~\ref{table1} shows the typical scales of the SUSY breaking parameters and the RH neutrino/sneutrino couplings in our model.
There are three questions that naturally arise given the parameters of our PeV scale SUSY breaking model in the table.
\begin{itemize}
\item What is the mass spectrum of the superpartners? In particular, which are the lightest and next-to-lightest superpartners?
\item Can we get electroweak symmetry breaking and the observed Higgs mass of 125 GeV with PeV scale SUSY breaking?
\item Given that the Yukawa couplings of the RH neutrinos are extremely suppressed, is it possible to obtain the right order-of-magnitude for the neutrino masses via the see-saw mechanism?
\end{itemize}
We will address these three issues in the remainder of this section.

\subsection{Mass hierarchy, sneutrino mass eigenstates and benchmark spectrum }
\label{sneutrinomassmatrix}
We assume in our model that the gravitino is the LSP after considering RG evolution of the SUSY breaking parameters. Since $R$-parity is conserved in this model, it is natural to then expect that the gravitino might exist as a viable DM component. We will assume that the lightest RH sneutrino is the NLSP. We will show in \textsection \ref{sect:lifetime} and \textsection \ref{relic} that the lightest RH sneutrino can also co-exist as another quasi-stable, long-lived DM component (albeit with a much smaller contribution to the relic density) because of its almost vanishing Yukawa and trilinear scalar couplings. In \textsection \ref{icecube}, we will show that the decays of the RH sneutrino can explain the high energy neutrino flux observed at IceCube.

We also assume that the mass hierarchy of the remaining superpartners follows the trend $M_a > m_{i} > m_{\tilde h_u} >  m_{N} > m_{\tilde g} > m_{\tilde N}  > m_{3/2}$. Here, $M_a$ is an electroweakino mass, $m_{i}$ is a slepton/squark mass parameter (other than RH sneutrino), $m_{\tilde h_u}$ is the up-type higgsino mass, $m_{N}$ is the RH neutrino mass, $m_{\tilde g}$ is the gluino mass, $m_{\tilde N}$ is the RH sneutrino mass\footnote{The RH sneutrino can be made lighter than the RH neutrino by choosing the sign of the relevant higher dimensional K\"{a}hler term which determines the mass splitting.} and $m_{3/2}$ is the gravitino mass. While the exact hierarchy of particles other than the gravitino and RH sneutrino is not absolutely critical for our main results, working with a specific hierarchy allows us to streamline our discussion in subsequent sections.

The masses of all the superpartners including the sneutrino are naturally on the order of the PeV scale. However, in the case of the sneutrino, the small trilinear scalar parameter $A_N$ leads to a tiny mixing between left-handed and right-handed sneutrinos. In order to calculate the physical eigenstates and their eigenvalues, we need to diagonalize the sneutrino mass matrix.
The full set of SUSY breaking parameters corresponding to the neutrino and other visible superfields (including all three matter generations) is given by
\bea
{\cal L}_\textrm{SUSY Br} \supset  &-& \tilde{Q}_L^\dagger m^2_{\tilde Q} \tilde{Q}_L
	- \tilde{u}_R^\dagger m^2_{\tilde u} \tilde{u}_R	- \tilde{d}_R^\dagger m^2_{\tilde d} \tilde{d}_R
	+ (- \tilde{u}_R^\dagger A_u \tilde{Q}_L\cdot h^0_u
          + \tilde{d}_R^\dagger A_d \tilde{Q}_L\cdot h^0_d + h.c.) \nonumber \\
	&-& \tilde{\ell}_L^\dagger m^2_{\tilde \ell} \tilde{\ell}_L
	- \tilde{N}_R^\dagger m^2_{\tilde N} \tilde{N}_R	- \tilde{e}_R^\dagger m^2_{\tilde e} \tilde{e}_R
	+ (- \tilde{N}_R^\dagger A_N \tilde{\ell}_L\cdot h^0_u
  + \tilde{e}_R^\dagger A_e \tilde{\ell}_L\cdot h^0_d + h.c.) \nonumber \\
	&+& \left[ (\tilde\ell\cdot h^0_u)^T \frac{c_{\ell}}{2} (\tilde\ell\cdot h^0_u)^* + \tilde{N}_R^T \frac{B_N }{2} \tilde{N}_R + h.c.\right] \nonumber \\
	&+& (B_\mu h^0_u\cdot h^0_d + h.c.) \ ,
\label{LSUSYBr.EQ}
\eea
where the couplings are $3 \times 3$ matrices in generation space.

The sneutrino mass matrix can be written in the following basis \cite{Gopalakrishna:2006kr},
 \bea
{\cal L}^{\tilde\nu}_{\textrm{mass}} &=& -\begin{pmatrix}\snuL^\dagger  \ & \sNR^\dagger \ & \snuL^T \ & \sNR^T \end{pmatrix}
 {\cal M}^2_{\snu} \begin{pmatrix}\snuL \cr \sNR \cr \snuL^* \cr \sNR^* \end{pmatrix}.
 \eea
In this basis, we find the mass matrix to be of the form
 \bea
{\cal M}^2_{\snu} &=& \frac{1}{2}
\begin{pmatrix}m_{LL}^2 & m^{2 \dagger }_{RL} & -v_u^2 c_\ell^\dagger & v_u m_N y^\dagger_N \cr
 m^{2}_{RL}  & m_{RR}^2 & v_u m^{T}_N y^*_N  & - B_N^\dagger \cr
- v_u^2 c_\ell & v_u m^{*}_N y^T_N   & m_{LL}^{2\, *} & m^{2 T}_{RL}  \cr
v_u m^{\dagger}_N y_N  & - B_N  & m^{2 *}_{RL}  & m_{RR}^{2\, *}
\end{pmatrix} \ ,
\label{massmatrix}
\eea
where
\beq
m_{LL}^2 = (m^2_{\tilde \ell} + v^2_u c_{\ell} + \Delta_\nu^2), \  m^{2}_{RL} = -\mu^* v_d y_N + v_u A^\dagger_N
 \textrm{ and}  \ m_{RR}^2 = m_{\tilde N}^2.
\eeq

Here, $\Delta_\nu^2 = (m_Z^2/2)\cos{2\beta}$ is the $D$-term contribution. We note that the term $v_u A_N$ in the mass matrix results in $\snuL \leftrightarrow \sNR$ mixing. Ignoring the phases present in Eq.~\ref{massmatrix} which can  lead to CP violating effects, we write  ${\tilde \nu_L}$ and ${\tilde N_R}$ in terms of real fields as
   \beq
    \tilde  \nu_L = \frac{\tilde \nu^\prime_1 + i \tilde \nu^\prime_2}{\sqrt{2}}, \ \ \ \tilde N_R = \frac{\tilde N^\prime_1 + i \tilde N^\prime_2}{\sqrt{2}}.
    \label{eq:realfields}
   \eeq
With this, the mass matrix reduces to block diagonal form as
\bea
&&  {\hskip -0.3in} {\cal L}^{\tilde\nu}_{\textrm{mass}} = -\frac{1}{2}
 \begin{pmatrix}\snu_1^{^\prime T} & \sN_1^{^\prime T} & \snu_2^{^\prime T} & \sN_2^{^\prime T}
\end{pmatrix} \nonumber\\
&& \begin{pmatrix}
  m_{LL}^2 -  v_u^2 c_\ell  & m^{2 T}_{RL} + v_u y^T_N m^*_N & 0 & 0 \cr
 m^{2}_{RL} + v_u y_N m^\dagger_N & m_{RR}^2 - B_N  & 0 & 0 \cr
0 & 0 & m_{LL}^2 +  v_u^2 c_\ell &m^{2 T}_{RL} - v_u y^T_N m^*_N   \cr
0 & 0 & m^{2}_{RL} - v_u y_N m^\dagger_N& m_{RR}^2 + B_N
\end{pmatrix}   \begin{pmatrix}\snu^\prime_1 \cr \sN^\prime_1 \cr \snu^\prime_2 \cr \sN^\prime_2
\end{pmatrix}. \nonumber
\eea

Diagonalizing the above matrix by performing a unitary transformation, we get the physical mass eigenstates,
\bea
\begin{pmatrix}
\snu_i \cr \sN_i
\end{pmatrix}
 &=&
\begin{pmatrix}
\cos{\theta^{\snu}_i} & -\sin{\theta^{\snu}_i} \cr \sin{\theta^{\snu}_i} &
\cos{\theta^{\snu}_i}
\end{pmatrix}
\begin{pmatrix}
\snu^\prime_i \cr \sN^\prime_i
\end{pmatrix}  .\
\label{eq:diagfield}
\eea
By considering a single generation of neutrino superfields and neglecting flavor mixings for simplicity, the mixing angle is given by
\bea
\label{eq:theta_nu}
\tan{2\theta^{\snu}_i} &=& \frac{2  (m^{2}_{RL} \pm v_u y_N m^\dagger_N ) }{(m_{LL}^2 \mp c_\ell v^2_u) - (m_{RR}^2 \mp B_N  )} \ , \label{thsnu.EQ}
\eea
with the top (bottom) sign for $i=1$ ($i=2$). Thus, the physical mass eigenstates of the sneutrino are given by $\sN_1$, $\sN_2$, ${\snu}_1$ and ${\snu}_2$.
We note that the mixing between left and right-handed sneutrinos is very small, since the value of the mixing angle based on the generic parameter values in Table~\ref{table1} can be estimated as $\theta_i^{\tilde \nu} \sim 10^{-19}$. In the next few sections we will show that the lightest eigenstate (presumed to be $\sN_1$) can exist as the appropriate decaying, sub-dominant DM component in this model.

In later sections, we will consider the benchmark spectrum: $M_a =10$~PeV, $m_{\tilde h_u} = 9.5$ PeV, $m_N = 8$~PeV, $m_{\tilde g} = 7.5$ PeV, $m_{\sN_1} = 6.5$~PeV, and $m_{3/2} = 1.5$~PeV when evaluating numerical results.

\subsection{EWSB, Higgs mass and PeV scale SUSY breaking}
\label{Higgsmass}
The Higgs potential for the two Higgs doublets takes on the form of the standard MSSM Higgs potential. Given that all the parameters in the Higgs potential are on the order of the PeV scale, it is possible to obtain the SM VEV of $v=246$~GeV  with fine-tuning~\cite{Giudice:2004tc,Wells:2004di}. In terms of the SUSY breaking parameters, this fine-tuning shows up as a restriction on the allowed region of parameter space. Thus, as one would expect, we can obtain successful EWSB at the cost of a little-hierarchy problem. 

In addition, if we require the light Higgs to have a mass of $\sim$ 125~GeV, this would lead to further restrictions on the space of allowed parameters. Below, we demonstrate the existence of a SM-like light Higgs boson for a suitable choice of SUSY breaking parameters.

The $2\times2$ mass-squared matrix for the neutral CP-even Higgs sector is given by~\cite{Djouadi:2005gj},
\bea
\label{eq:Hmatrix_loop}
{\cal M}^2= \begin{pmatrix}
{\cal M}^{2}_{11} + \Delta {\cal M}^{2}_{11} ~~&~~  {\cal M}^{2}_{12} + \Delta{\cal M}^{2}_{12} \cr
{\cal M}^{2}_{12} + \Delta{\cal M}^{2}_{12}  ~ ~& ~~{\cal M}^{2}_{22} + \Delta{\cal M}^{2}_{22}
\end{pmatrix},
\eea
where the entries ${\cal M}^{2}_{ij}$ correspond to the tree level masses of the neutral CP-even Higgs doublets and the terms $ \Delta {\cal M}^{2}_{ij}$ take into account the loop corrections. The dominant corrections arise from one-loop corrections which are controlled by the top/stop couplings, subleading one-loop corrections which are controlled by bottom/sbottom couplings and two-loop corrections due to the top/bottom Yukawa couplings and the strong coupling constant. The tree level mass-squared matrix for the neutral CP-even Higgs sector is given by
\bea
{\cal M}^{2}_{\rm tree}= \begin{pmatrix}
{\cal M}^{2}_{11} ~ & ~ {\cal M}^{2}_{12} \cr
{\cal M}^{2}_{12} ~ &~ {\cal M}^{2}_{22} \nonumber\\
\end{pmatrix}=
\begin{pmatrix}
M^2_A \sin^2{\beta} + M^2_Z \cos^2{\beta}  ~ & ~ -\left(M^2_A + M^2_Z \right) \sin{\beta}\cos{\beta} \cr
-\left(M^2_A + M^2_Z\right) \sin{\beta}\cos{\beta} ~ &~ M^2_A \cos^2{\beta} + M^2_Z \sin^2{\beta}
\end{pmatrix},
\eea
and
\bea
&& \Delta {\cal M}^{2}_{11} = -\frac{v^2 \sin^2\beta}{32\pi^2}{\bar \mu}^2 \Bigl[x^2_t \lambda^4_t (1+ c_{11} l_s)+a^2_b \lambda^4_b (1+c_{12}) l_s\Bigr], \nonumber\\
&& \Delta {\cal M}^{2}_{12} = -\frac{v^2 \sin^2\beta}{32\pi^2}{\bar \mu} \Bigl[x_t \lambda^4_t (6 - x_t a_t)(1+ c_{31} l_s) - {\bar \mu}^2 a_b \lambda^4_b (1+c_{32} l_s)\Bigr],\\
&& \Delta {\cal M}^{2}_{22} = \frac{v^2 \sin^2\beta}{32\pi^2} \Bigl[6 \lambda^4_t l_s(2+ c_{21} l_s) + x_t a_t \lambda^4_t (12- x_t a_t)(1+ c_{21} l_s)- {\bar \mu}^4 \lambda^4_b (1+c_{22}l_s) \Bigr], \nonumber
\eea
where $m_Z$ is the $Z$-boson mass, $\tan \beta$ is the ratio of up-type to down-type Higgs VEVs, $m_A$ is the mass of the CP-odd Higgs and $\lambda_{t,b}$ are the top and bottom Yukawa couplings respectively. The shorthand notation used in these formulas is to be interpreted as follows:  $l_s= {\rm Log}[{M^2_S}/{m^2_t}]$, ${\bar \mu}=\mu/M_S$, $a_{t,b}= A_{t,b}/M_S$ and $x_t=X_t/M_S$. Here, $M_S$ is the arithmetic mean of the stop squark masses, i.e. $M_{S}= \frac{1}{2}({ m_{{\tilde t}_1}} + { m_{{\tilde t}_2}})$, $m_t$ is the top quark mass, $A_{t,b}$ are the stop and sbottom trilinear couplings respectively, and  $X_t=A_t- \mu \cot\beta$. The coefficients $c_{ij}$ correspond to the leading two-loop corrections due to the top/bottom Yukawa couplings and strong coupling constant $g_3$; they are given by,
 \bea
 c_{ij}=\frac{1}{32 \pi^2}\left(t_{ij} \lambda^2_t + b_{ij}\lambda^2_b - 32 g^2_3\right),
 \eea
 with $$(t_{11},t_{12},t_{21},t_{22},t_{31},t_{32})=(12,-4,6,-10,9,7),$$
 $$(b_{11},b_{12},b_{21},b_{22},b_{31},b_{32})=(-4,12,2,6,18,-1).$$
\begin{figure}
\begin{center}
\includegraphics[width =  0.8\textwidth]{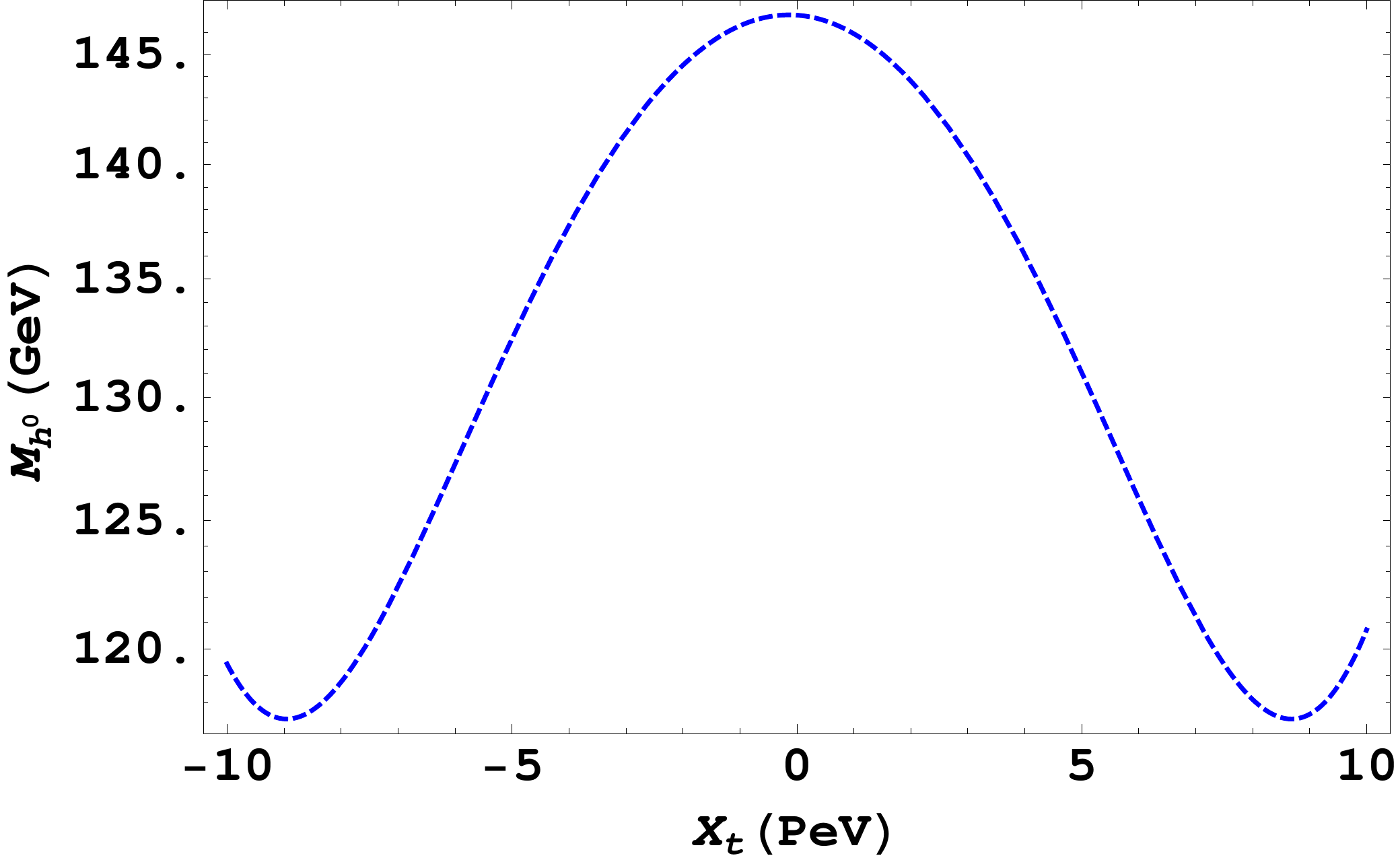}
\caption {The figure shows the radiatively corrected light Higgs mass as a function of $X_t$ for $M_A=8$~PeV, $M_S=3.6$~PeV, $\mu=9.5$~PeV, $A_b=90$~PeV and $\tan\beta \sim 50$. We can see that for $X_t=\pm6$~PeV, we can obtain a light Higgs boson of mass $\sim$ 125 GeV. Thus, by fine-tuning the masses of PeV scale superpartners and other soft SUSY breaking parameters, it is possible to obtain a SM-like light Higgs boson.}
\label{Higgs_Mass}
\end{center}
\end{figure}
The mass of the physical Higgs bosons can be obtained by diagonalizing the Higgs mass matrix as given in Eq. \ref{eq:Hmatrix_loop}. 
We have fixed a benchmark set of parameters $M_A=8$~PeV, $M_S=3.6$~PeV, $\mu=9.5$~PeV, $A_b=90$~PeV, and $\tan\beta \sim 50$ and computed the eigenvalues of the mass matrix for different values of $X_t$. In Fig.~\ref{Higgs_Mass}, we have shown the smaller eigenvalue (which corresponds to the mass of the light Higgs) as a function of $X_t$, for this choice of parameters. We find that for $X_t=\pm6$~PeV, we can obtain a light Higgs boson of mass $\sim$ 125 GeV. For this set of parameters, the heavy Higgs boson has a mass of 8 PeV. Thus, we have demonstrated that a light Higgs boson can easily be obtained in this set up.

Note that since the superpartners and the other Higgs bosons are very heavy, the light Higgs boson has very nearly Standard Model like couplings. A full scan over the parameter space to find the phenomenologically acceptable SUSY breaking parameters is beyond the scope of this work, but these can be found by using numerical codes such as SUSPECT \cite{Djouadi:2002ze} and SOFTSUSY \cite{Allanach:2001kg}.

\subsection{Neutrino masses}
\label{neutrino}
We would also like to show that our model explains the small value of the observed neutrino masses despite the suppressed neutrino Yukawa coupling $y_N$. In generic models with RH neutrinos, the masses of the neutrinos are controlled by the following mass matrix which mixes left and right-handed neutrinos:
\bea
m_{i}= \begin{pmatrix}
m_{\nu_L} &   y_N v_u \cr
y_N v_u   & M_N
\end{pmatrix}.
\eea
Here $m_{\nu_L}$ and $M_N$ are Majorana like masses for the left and right handed neutrinos and $v_u y_N$ is a Dirac mixing induced by a VEV for the up-type MSSM Higgs. In the conventional see-saw mechanism, $m_{\nu_L}$ is either zero or suppressed. With this, the lightest eigenvalue of the neutrino mass matrix is given by $y^2_N v^2_u /M_N$, with $M_N$ being the mass of the (heavy) RH neutrino. This mechanism for neutrino mass generation does not work in our model because of the extremely suppressed Yukawa coupling $y_N$. However in the context of supersymmetric models, various alternatives to the conventional see-saw mechanism have been conjectured to originate from different non-renormalizable K\"{a}hler operators present in the theory \cite{Frere:1999uv,ArkaniHamed:2000bq,ArkaniHamed:2000kj,Casas:2002sn,Brignole:2010nh}. In our model, the dominant source of neutrino masses arises from the following non-renormalizable K\"{a}hler operator present in the theory (see Eq.~\ref{eq:K}):
\beq
\int d^2 \theta d^2 {\bar \theta} \frac{1}{M^2_*} \left(\widehat L \cdot \widehat  H^{\dagger}_d \right) (\widehat  L\cdot  \widehat  H_u). \eeq
For $K \supset  \widehat H^{\dagger}_d \widehat  H_d$ and $W \sim m_{3/2} M^2_{\textrm{p}}$ (an explanation for expecting this is given in appendix {\bf A}),  we get $F_{H_d} \sim \langle h^0_d \rangle m_{3/2}$, where $\langle h^0_d \rangle = v \cos \beta$ is the VEV of the down-type Higgs. By expanding $\widehat L = l + \theta {\tilde l}$ and setting  $h^0_{u,d}$ to their VEVs, this gives
\beq
m_{\nu_L} \sim \frac{m_{3/2} v^2 \sin\beta \cos\beta}{  M^{2}_{*}}.
\eeq

For $v=246$ GeV, $\tan\beta\sim 1$ and $M_{*} \sim 10^{10}$ GeV, we get $m_{\nu} \sim 0.1$ eV. Since the off-diagonal term in the neutrino mass-matrix is almost vanishing, we do not expect any large additional contributions to the light neutrino masses from the see-saw with the RH neutrino $N$. Thus, upon diagonalization of the neutrino mass matrix we get a purely Majorana, light neutrino with mass $m_1= m_{\nu_L} \sim 0.1$ eV and a heavy (approximately) right-handed neutrino with mass $m_2 = m_{N}= {\cal O}(1)$ PeV.

Here, we have introduced another scale $M_{*}$ based on the assumption that the model exhibits another symmetry at this intermediate scale. Our implicit assumption is that all the visible supersymmetric fields have some non-zero charges under this new symmetry while the supersymmetry breaking field (the hidden-sector field) is uncharged under this symmetry\footnote{This particular assumption
has been also considered in ref.~\cite{Brignole:2010nh}. In that work, the authors have argued that the non-renormalizable operators involving all the supersymmetric fields are suppressed by a scale $M$ whereas the non-renormalizable operators involving the
supersymmetry breaking fields are suppressed by a scale $M_S$ (where $M_S$ could be either larger or smaller than $M$).}. Therefore, any non-renormalizable interaction term involving only the visible sector superfields (and in particular the term $ (\widehat L \cdot \widehat  H^{\dagger}_d ) (\widehat  L\cdot  \widehat  H_u)$), will be suppressed by $M_{*}$ instead of $M_p$. On the other hand, all the non-renormalizable interactions involving the hidden-sector field (for e.g. the term that generates the neutrino Yukawa coupling in our model, $   \widehat X^\dagger  \widehat N^c \widehat L \cdot\widehat H_u$) are still suppressed by the Planck scale.

\section{Decay lifetime of the RH sneutrino}
 \label{sect:lifetime}

\begin{figure}
\begin{center}
\includegraphics[width =  \textwidth]{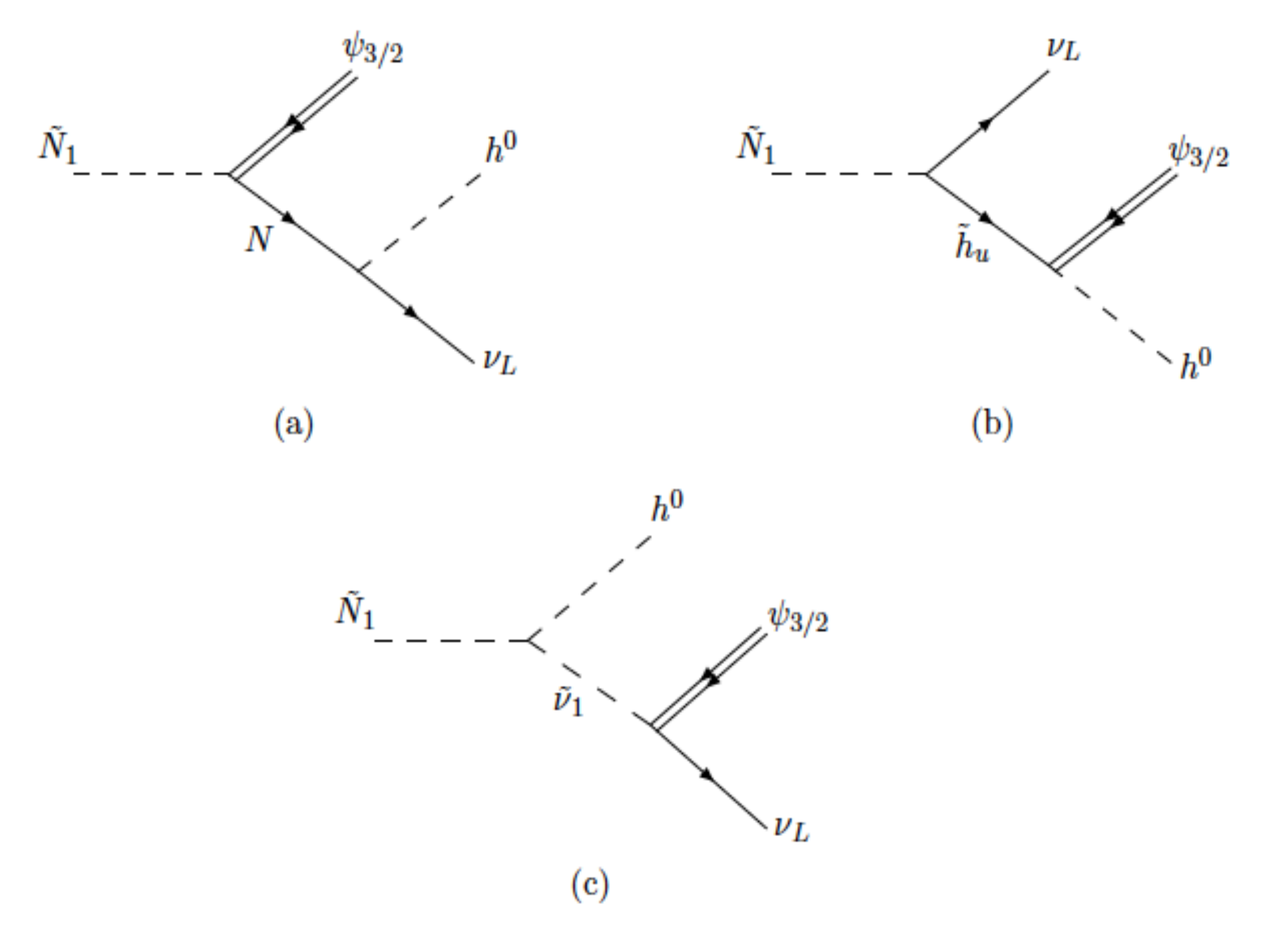}
\caption {Feynman diagrams corresponding to the decay mode $ \sN_1 \rightarrow \nu_L h^0 \psi_{3/2}$. The diagrams (a) and (b) show interactions which proceed through the Yukawa interaction with coupling $y_N$, whereas diagram (c) proceeds through the trilinear interaction with strength proportional to $A_N$. The Yukawa interaction dominates the decay process, and in fact leads to the three-body decay being dominant over the two-body decay $ {\sN_1}\rightarrow \nu_L  {\psi_{3/2}}$ which is suppressed by the small mixing angle.}
\label{feyn_decay}
\end{center}
\end{figure}

In this section, we will calculate the decay width of the lightest right-handed sneutrino.
Due to $R$-parity conservation and the assumed mass hierarchy: $M_a > m_{i} > m_{\tilde h_u} >  m_{N} > m_{\tilde g} > m_{\tilde N}  > m_{3/2}$, the possible decay modes of the sneutrino include:

\begin{enumerate}
\item[(a)]
The two-body decay $ {\sN_1}\rightarrow \nu_L  {\psi_{3/2}}$, which occurs through mixing with the LH sneutrino.
\item[(b)]
The three-body decay  $ \sN_1 \rightarrow  {\psi_{3/2}} \nu_L  h^0$. The Feynman diagrams for this decay mode are shown in Fig.~\ref{feyn_decay}\footnote{Other suppressed decay modes through mixing are the three body decays:  ${\sN_1}\rightarrow \nu_L  {\tilde B}^*$,     ${\sN_1}\rightarrow e^{\pm}_L  {\tilde W^{*\mp}}$, where the off-shell gauginos ${\tilde B}^*/{\tilde W^{*\mp}}$ convert into $ \psi_{3/2}$ and the corresponding SM gauge bosons. These decay widths are extremely suppressed by both three body phase space as well as the small mixing angle.}.
\end{enumerate}

We note that both these decay modes are suppressed. The first decay mode is parameterically suppressed through the small mixing between left-handed and right-handed sneutrinos (which is given by $\sin\theta_i^{\tilde \nu}$) while the second decay mode (depending on the relevant Feynman diagram) is suppressed either by the small Yukawa coupling ($y_N$) or by the trilinear interaction coupling ($A_N$). Of all these suppressed couplings, the Yukawa coupling turns out to be the least suppressed, and thus the decay of the RH sneutrino is dominantly to the three body final state $ \sN_1 \rightarrow  {\psi_{3/2}} \nu_L  h^0$. We will justify this below.

We will begin by estimating the partial decay widths. Since all the decay channels involve a gravitino as one of the final decay products, we first write the relevant gravitino ($\psi_\mu$) interaction terms in the Lagrangian which are given by \cite{Wess:1992cp},
\bea
\label{eq.gravitino_int}
&& {\cal L} \sim -{\frac{i}{ 2 M_{\textrm{p}}}} \left[ (D_\mu^* \phi^*) {\bar \psi_\nu}  \gamma^\mu \gamma^\nu P_L \chi^i -  (D_\mu \phi) {\bar \psi_\nu}  \gamma^\mu \gamma^\nu P_R \chi^i   \right]   -{\frac{i}{ 8  M_{\textrm{p}}}} {\bar \psi_\nu} [ \gamma^\nu, \gamma^\lambda] \gamma^\mu \lambda^{\alpha(a)} F^{\alpha(a)}_{\nu \rho}, \nonumber\\
\eea
where $\chi_i$ stands for any chiral left-handed Weyl spinor,  $\phi_i$ stands for the scalar partner of $\chi_i$ in the same chiral supermultiplet, $\lambda^{\alpha(a)}$ and  $F^{\alpha(a)}_{\nu \rho}$ denote the gaugino and gauge field strength in the same vector supermultiplet. At high energies \cite{Casalbuoni:1988kv,Maroto:1999vd}, the gravitino field can be replaced by the goldstino by a generalization of the Goldstone boson equivalence principle, $\psi_\mu  \rightarrow \frac{ D_\mu \chi}{\sqrt{3} m_{3/2}}$.

The non-zero mixing between the left-handed and right-handed sneutrino allows for two-body decays of the type ${\sN_1}\rightarrow \nu_L  {\psi_{3/2}}$.  Using the interaction terms given in Eq.~\ref{eq.gravitino_int}, we can estimate the partial decay width for this mode as,
\bea
\Gamma_{\sN_1 \rightarrow \nu_L  \psi_{3/2}} &\sim \frac{1}{48 \pi} \frac{m^5_{\sN_1} }{m^2_{3/2} M^2_{\textrm{p}}}  \sin^2\theta_i^{\tilde \nu}, \\
                                            &\sim \frac{1}{48 \pi} \frac{m^3_{\rm PeV}  }{ M^2_{\textrm{p}}}\sin^2\theta_i^{\tilde \nu}, \\
                                            &\sim 10^{-58} \textrm{ GeV}.\label{eq:neutrino2body}
\eea
where we have used $m_{\rm PeV}$ as a generic representative SUSY breaking PeV mass and in the numerical expression we have used the estimated value of $\sin \theta_i^{\tilde \nu} \sim 10^{-19}$.

The three body decay mode $ \sN_1 \rightarrow  {\psi_{3/2}} \nu_L  h^0$ is generally expected to be phase space suppressed relative to the two-body decay. However, looking at the Feynman diagrams for this decay mode, we can see from Fig.~\ref{feyn_decay} (a) and (b), that the decay proceeds through the Yukawa interaction $y_N \sim 10^{-13}$, which is parameterically much larger than the mixing factor $\sin \theta_i^{\tilde \nu}$\footnote{We note that the Yukawa coupling is also much larger than $A_N/m_{\rm{PeV}}$ and thus the Feynman diagram of Fig.~\ref{feyn_decay} (c) is also parameterically suppressed.}.  The decay width for this three body mode is therefore given by:
\bea
\Gamma_{\sN_1 \rightarrow \nu_L h^0  \psi_{3/2}}  &\sim \frac{1}{2^{9}. 9 \pi^3}  \frac {y_N^2 m^3_{\rm PeV}}{M_{\textrm{p}}^2}, \\
                                                   &\sim 10^{-50}~{\rm GeV}. \label{eq:neutrino3body}
\eea
Comparing Eqs.~(\ref{eq:neutrino2body}) and (\ref{eq:neutrino3body}), we therefore see that the three-body decay is much more dominant than the two-body decay decay mode.

Thus, the lifetime of the right-handed sneutrino turns out to be,
\beq
\tau_{\sN_1} = \frac{\hbar}{\Gamma_{\sN_1 \rightarrow \nu_L h^0 \psi_{3/2}}} \sim 10^{26} {\textrm{ sec}}. \label{eq:snlifetime}
\eeq
Since the lifetime is longer than the age of the universe, the RH sneutrino can act as a quasi-stable DM component co-existing with the stable gravitino DM. As it is evident that the sneutrino decays gives rise to neutrinos, the late decay of a PeV-scale sneutrino could potentially explain the flux of PeV scale neutrino events observed at IceCube. We will discuss this issue in \textsection \ref{icecube}.

\section{Relic abundance}
\label{relic}
In the previous section, we found that in addition to the gravitino LSP, the real component of the lighter right-handed sneutrino ($ \sN_1$) can co-exist as a viable DM component in our model. Before we explore whether sneutrino decays can explain the IceCube PeV events, we compute the relic abundance of both the proposed DM components. Since both components have extremely suppressed interactions with the SM particles, they would not be in thermal equilibrium in the early universe. However, a non-zero relic density for both particle species can be produced from the decay or inelastic scattering of heavier superpartners existing in thermal equilibrium with the SM plasma. This process is known as freeze-in production of DM. In order to keep the discussion self-contained, we briefly review the details of the freeze-in mechanism from decay and scattering of heavier particles~\cite{Hall:2009bx,Elahi:2014fsa}. We will consider below the two generic processes: $\phi \rightarrow \psi + \chi$ and $\phi_1 + \phi_2 \rightarrow \psi + \chi$, with $\chi$ being the weakly-coupled DM component.

We first discuss the production of $\chi$ from the decays of a heavier species $\phi$.
The evolution of the number density of $\chi$ is described by the following Boltzmann equation~\cite{Hall:2009bx,Elahi:2014fsa}:
\bea
&& \dot{n}_{\chi} + 3 H n_{\chi} = \int d\Pi_{\phi} d\Pi_{\psi} d\Pi_{\chi} (2 \pi)^4 \delta^4(p_{\phi} + p_{\psi} -p_{\chi}) \nonumber\\
&& \times \left[ \left|\cal M\right|^2_{\phi \rightarrow \psi + \chi} f_{\phi} (1 \pm f_{\psi}) (1 \pm f_{\chi}) - \left|\cal M\right|^2_{\psi + \chi \rightarrow \phi} f_{\psi} f_{\chi} (1\pm  f_{\phi}) \right],
\eea
where the factors of $d\Pi_{\phi_j}= d^3 p_j/( (2\pi)^3 2 E_j)$ correspond to phase space measure and the factors of $f_j$ corresponds to phase space density of particles of type $j$. The sum over the initial and final spins in the squared matrix elements is assumed implicitly in this equation. Assuming that the initial abundance of $\chi$ is zero, we will have $f_{\chi}=0$. Thus we can neglect Pauli blocking and Bose enhancement effects, and set the factors of $(1\pm f_{\chi}) \approx 1$. Performing the two body phase space integration over the kinematic distributions of $\psi$ and $\chi$ and writing the result in terms of the decay width $(\Gamma_{\phi})$ and mass $(m_{\phi})$ of $\phi$, the equation above simply reduces to
\bea
&& \dot{n}_{\chi} + 3 H n_{\chi} = g_{\phi} \int d\Pi_{\phi}  \Gamma_{\phi} m_{\phi} f_{\phi} = g_{\phi} \int \frac{d^3p_{\phi}}{(2 \pi)^3}  \Gamma_{\phi} m_{\phi} f_{\phi},
\eea
where $g_{\phi}$ is the spin degeneracy of $\phi$. Since the $\phi$ particles are assumed to be dilute and in thermal equilibrium, we can approximate $f_{\phi} \simeq e^{-E_{\phi}/T}$, where $T$ is the temperature of the kinetically-coupled $\phi$ particles. Changing the integral over momentum space into an integral over energy, one obtains
\bea
&& \dot{n}_{\chi} + 3 H n_{\chi} = g_{\phi} \int \frac{m_{\phi} \Gamma_{\phi}}{2 \pi^2} (E^2_{\phi} - m^2_{\phi})^{1/2}  e^{-E_{\phi}/T}  d E_{\phi} =  \frac{g_{\phi} m^2_{\phi} \Gamma_{\phi}}{2 \pi^2} T K_{1}(m_{\phi}/T),
\eea
where $ K_{1}$ corresponds to the first Bessel function of the second kind. Writing $n_\chi$ in term of the yield $Y_\chi = n_{\chi}/S$ (where $S$ is the entropy density of the universe at a particular temperature $T$) and utilizing $\dot{T} \sim - HT$, we obtain
\bea
Y_{\chi(\rm decay)} \approx \int^{T_\textrm{max}}_{T_\textrm{min}} \frac{g_{\phi} m^2_{\phi} \Gamma_{\phi}}{2 \pi^2} \frac{ K_{1}(m_{\phi}/T)}{ S(T) H(T)} dT,
\eea
where the entropy density, $S(T)$ and the Hubble rate, $H(T)$ are given by $S(T)= 2 \pi^2 g^S_{*} T^3/45$ and $H(T)= 1.66 \sqrt{g^{\rho}_{*}} T^2/M_{\textrm{p}}$ in the radiation dominated era, and $g^S_{*}$ and $g^{\rho}_{*}$ are the effective entropy and energy degrees-of-freedom present in the radiation bath. After defining $x =m_{\phi}/T$, the integral turns out to be
\bea
Y_{\chi(\rm decay)} \approx  \frac{45 \ g_{\phi}  \Gamma_{\phi} M_{\textrm{p}} }{(1.66) \ 4\pi^4 \ m^2_{\phi} \  g^S_{*} \sqrt{g^{\rho}}} \int^{x_{\rm max}}_{x_{\rm min}} K_{1}(x) x^3 dx.
\eea
Approximating the limits of integration as ${x_{\rm max}} = \infty$ and ${x_{\rm min}} =0$, we get $\int^{x_{\rm max}}_{x_{\rm min}} K_{1}(x) x^3 dx = 3 \pi/2$. Using this, the final expression for $Y^{\chi}_{\phi \rightarrow \psi \chi}$  turns out to be
\bea
\label{eq:decay}
Y_{\chi(\rm decay)} \approx  \frac{135 \ g_{\phi}}{ (1.66) \ 8 \pi^3 g^S_{*} \sqrt{g^{\rho}}} \left(\frac{M_{\textrm{p}} \Gamma_{\phi}}{m^2_{\phi}}\right).
\eea

Similarly we can calculate $Y_{\chi}$ for the scattering process $\phi_1 + \psi \rightarrow \phi_2 + \chi$. In particular, we discuss the case where the scattering process involves non-renormalizable operators i.e. of the form ${\cal L} \supset \frac{1}{\Lambda} \phi_1 { \psi} {\bar{ \chi}} \bar{\phi}_2$. At temperatures much higher than the masses of all particles involved, the squared-matrix amplitude for the given scattering process is proportional to,
\beq
\left|{\cal M} \right|^2 \sim \kappa \frac{s}{\Lambda^2},
\eeq
where $s$ is the interaction center of mass (c.o.m.) energy squared at a temperature $T$. The value of $\kappa$ is determined by other low energy parameters such as masses of the particles involved in the scattering process. The change in number density of $\chi$ due to this process is described by the following Boltzmann equation:
\bea
&& \dot{n}_{\chi} + 3 H n_{\chi} = \int d\Pi_{\phi_1} d\Pi_{\phi_2} d\Pi_{\psi} d\Pi_{\chi} (2 \pi)^4 \delta^4(p_{\phi_1} + p_{\psi} - p_{\phi_2} -p_{\chi}) \nonumber\\
&& \times \left[ \left|\cal M\right|^2_{\phi_1 \psi \rightarrow \phi_2 + \chi} f_{\phi_1} f_{\psi} (1 \pm f_{\phi_2}) (1 \pm f_{\chi}) - \left|\cal M\right|^2_{\phi_2 + \chi \rightarrow \phi_1 \psi } f_{\phi_2} f_{\chi} (1\pm  f_{\phi_1})  (1\pm  f_{\psi}) \right].
\eea
Assuming that the initial abundance of $\chi$ is negligible, we set $f_{\chi} =0$. Following \cite{Elahi:2014fsa}, the collision term can be written as an integral with respect to the c.o.m. energy:
\bea
&& \dot{n}_{\chi} + 3 H n_{\chi} =  \frac{ 3 T}{ 512 \pi^6} \int ds  \ d \Omega \  P_{\phi_1 \psi}P_{\phi_2 \chi} \left|{\cal M}\right|^2 \frac{1}{\sqrt{s}} K_1 \left(\frac{\sqrt{s}}{T}\right),
\eea
with
\beq
P_{ij} = \frac{1}{2 \sqrt{s}} \sqrt{s - (m_i + m_j)^2} \sqrt{s - (m_i - m_j)^2}.
\eeq
 For $\left|{\cal M} \right|^2 \sim \kappa \frac{s}{\Lambda^2}$, the equation above reduces to
 \bea
&& \dot{n}_{\chi} + 3 H n_{\chi} =  \frac{\kappa \cdot T}{ 512 \pi^5 \Lambda^2} \int ds \  s^{3/2} K_1 \left(\frac{\sqrt{s}}{T}\right) \approx \frac{\kappa \cdot T^6}{16 \pi^5 \Lambda^2}.
\eea
Expressing  $n_\chi$ in terms of the yield $Y_\chi=n_\chi/S$, we can rewrite the Boltzmann equation as:
\beq
\frac{ dY_{\chi(\rm scattering)}}{dT} \approx -\frac{1}{ S H T} \frac{\kappa \ T^6}{ 16 \pi^5 \Lambda^2} \approx \frac{ 45\  \kappa \ M_{\textrm{p}}}{ (1.66) \ 32\pi^7  g^S_{*} \sqrt{g^{\rho}} \Lambda^2}.
\eeq
Integrating this with respect to temperature by using  limit of integrations to be $T_{\min}=0$ and $T_{\max} = T_R$ (the reheating temperature), the final expression for $Y^{\chi}_{\phi_1  \psi \rightarrow \phi_2  \chi}$ is just given by
\beq
\label{eq:scattering}
Y_{\chi(\rm scattering)} \approx \frac{180}{ 1.66 \ (2 \pi)^7 g^S_{*} \sqrt{g^{\rho}}} \left(\frac{\kappa \ T_{R} M_{p}}{\Lambda^2}\right),
\eeq
indicating that the number density produced through the scattering process is linearly proportional to the reheating temperature ($T_R$).

Next we will use the above results to calculate the relic abundance of both the gravitino as well as the RH sneutrino.

\paragraph{Relic abundance of the gravitinos:} Since the heavy gravitinos decouple from the thermal plasma in the very early universe because of their Planck suppressed interactions, they do not have a significant population in the early universe from thermal production. The dominant number density of heavy gravitinos will therefore be produced from the decay and/or inelastic scattering of heavier particles present in the early universe. Utilizing the gravitino interaction terms given in Eq.~\ref{eq.gravitino_int}, the possible processes involving the decay of heavier super-partners $(i)$ into the gravitino include
\beq
{i} \rightarrow  \psi_{3/2} + {\rm SM}.
\eeq

In the high energy limit, when $\psi_{3/2 (\mu)} \rightarrow {\partial_\mu \chi}/m_{3/2}$, the decay width of any heavier particle decaying into a gravitino is given by
\bea
&& \Gamma_{(i \rightarrow \psi_{3/2} + ..)} \sim \frac{1}{48 \pi } \frac{m^5_i}{m^2_{3/2} M^2_{\textrm{p}}}, \eea
where $m_i$ corresponds to the mass of the heavier superpartner.  Using Eq. \ref{eq:decay}, the gravitino yield $Y_{3/2 (\rm decay)}$ will be given by
\bea
Y_{3/2 (\rm decay)} &\approx&  \frac{45 \ g_a}{(1.66) \ 2^7 \pi^4  g^S_{*} \sqrt{g^{\rho}}} \left(\frac{ M^3_{a} }{m^2_{3/2} M_{\textrm{p}}}\right), \\
 &\approx& 2.8 \times 10^{-16} \left( \frac{ M_{a} }{10 \textrm{ PeV}} \right )^3  \left(\frac{1.5 \textrm{ PeV}}{m_{3/2}} \right)^2, \label{eq:decay_2}
\eea
where we have made the simplifying assumption that electroweakinos are the heaviest sparticles and their decays dominate the production of gravitinos compared to other sparticle decays. Note that we have taken $g^S_{*} \sqrt{g^{\rho}} \sim (200)^{3/2}$ in the final expression.

Similarly there are many inelastic scattering processes that contribute to the production of the gravitino. The dominant scattering processes proceed through non-renormalizable operators e.g. production of gravitinos from the process: ${q} + {\tilde g} \rightarrow \psi_{3/2} + { q}$, where $q$ is a SM quark and $\tilde g$ is the gluino. The complete list of possible scattering processes is given in \cite{Bolz:2000fu} and the matrix amplitude is given by
\beq
\left|{\cal M}\right|^2_{(i + j  \rightarrow \psi_{3/2} + ..)} \propto \frac{1}{M^{2}_p} \left( 1+ \frac{m^{2}_{\tilde g}}{3 m^{2}_{3/2}}\right),
\eeq
where $m_{\tilde g}$ corresponds to the mass of gluino.

Further using Eq.~\ref{eq:scattering}, and using $\kappa \sim  (1+ {m^{2}_{\tilde g}}/{3 m^{2}_{3/2}}) \sim {m^{2}_{\tilde g}}/{(3 m^{2}_{3/2})}$ and $\Lambda= M_{\textrm{p}}$, the gravitino yield from inelastic scattering of the gluino $Y_{3/2(\rm scattering)}$ is given by
\bea
Y_{3/2(\rm scattering)} &\approx& \frac{60}{ 1.66 \  (2 \pi)^7 g^S_{*} \sqrt{g^{\rho}}} \left(\frac{m^{2}_{\tilde g} \ T_{R} }{m^{2}_{3/2} M_{\textrm{p}}}\right), \\
  &\approx& 3.3 \times 10^{-18} \left(\frac{m_{\tilde g}}{7.5 \textrm{ PeV}}  \right)^2   \left(\frac{1.5 \textrm{ PeV}}{m_{3/2}} \right)^2 \left( \frac{T_{R}}{10 \textrm{ PeV}} \right ) . \label{eq:scattering_2}
\eea

The present day relic abundance of any species $\chi$ is related to the yield by \cite{Hall:2009bx}:
\beq
\label{eq:relic}
\Omega_{\chi}  = \frac{m_\chi  Y_\chi  S(T_0)}{\rho_c},
\eeq
where the present day entropy density, $S(T_0) = 2 \pi^2 g^S_{*} {T^3_0}/45 \sim (2.8 \times 10^{-4}~{\rm eV})^3$ for the current CMB temperature of $ T_0 = 2 \times 10^{-4}$ eV and the critical density is $\rho_c/h^2 = (2. 95 \times 10^{-3} {\rm eV})^4$.

As is evident from Eq.~\ref{eq:decay_2} and Eq.~\ref{eq:scattering_2}, for a reheating temperature not much larger than the PeV scale, both scattering and decay processes contribute comparably to the gravitino relic abundance via freeze-in. For our benchmark mass parameters $m_{3/2} =1.5$ PeV, $m_{\tilde g}=7.5$ PeV, $M_{a} = 10$ PeV and assuming $T_R \lesssim 1000$ PeV, electroweakino decay processes dominate over scattering processes and therefore determine the final yield of gravitinos.

The gravitino relic abundance can be estimated from Eq.~\ref{eq:decay_2} and Eq.~\ref{eq:relic} as
\beq
\Omega_{3/2} h^2  \sim  0.12 \left( \frac{ M_{a} }{10 \textrm{ PeV}} \right )^3  \left(\frac{1.5 \textrm{ PeV}}{m_{3/2}} \right).
\eeq
This is the first order-of-magnitude miracle that we promised. For superpartners masses near the PeV scale and a reheating temperature less than $\sim$~1000 PeV, we automatically get a relic abundance of gravitinos consistent with the observed amount of dark matter in the universe.

\paragraph{Relic abundance of sneutrinos:} As we have shown in \textsection \ref{sect:outline}, the right-handed sneutrino ($ \sN_1$) has very suppressed interactions with other MSSM particles. Therefore, as in the case of the gravitino, it would also never attain equilibrium with the thermal plasma. Similar to the gravitino abundance, the relic density of the sneutrino can be produced via decay and scattering of thermally produced heavy MSSM superpartners. However, since the production via scatterings proceeds through renormalizable operators, there is no reheating temperature enhancement and this contribution is always subdominant as compared to decay processes~\cite{Asaka:2005cn}. We will therefore neglect the contribution to RH sneutrino production from the scattering processes.
Below we list the possible two-body decay processes that might give a non-zero relic density for the sneutrino ($\sN_1)$:
\begin{enumerate}
\item$\tilde h_u ~ (\rm Higgsino)$  $\rightarrow \sN_1  \nu_L$ (mediated by Yukawa interactions),
\item  $\tilde \nu_1 ~(\rm Heavier~sneutrino)$  $\rightarrow \sN_1 h^0 $ (mediated by trilinear scalar interactions),
\item $  \tilde{e}_L  ~ (\rm Slepton)$  $\rightarrow \sN_1 W^{\pm} $ (mediated by the off-diagonal mixing angle and gauge interactions).
\end{enumerate}

As we have seen before, the Yukawa coupling is the dominant interaction coupling of the sneutrino and it determines the production rate of sneutrinos from the decay of a heavier species. The right-handed sneutrinos are therefore dominantly produced through the decay of $\tilde {h}_u$  $\rightarrow \sN_1  \nu_L $ which is mediated by the Yukawa interaction, and whose rate is given by
\beq
\Gamma_{{\tilde h_u}  \rightarrow \sN_1 \nu_L } \sim  \frac{y^{2}_N m_{\tilde h_u}}{32 \pi}.
\eeq
Incorporating this in Eq.~\ref{eq:decay}, the sneutrino yield $Y_{\sN_1({\rm decay})}$ is given by
\bea
Y_{\sN_1({\rm decay})} &\approx&  \frac{135 \ g_{\tilde h_u}}{  2^8 \pi^4 ~(1.66)~g^S_{*} \sqrt{g^{\rho}}} \left(\frac{ y^2_N M_{\textrm{p}}}{m_{\tilde h_u}}\right), \\
  &\approx& 7.5 \times 10^{-22} \left(\frac{ y_N}{0.5 \times 10^{-13}} \right )^2 \left( \frac{9.5 \textrm{ PeV}}{m_{\tilde h_u}} \right ).
\eea
Plugging in this expression for $Y_{\sN_1({\rm decay})}$  in Eq.~\ref{eq:relic}, the relic abundance of the sneutrino ($\sN_1$) is given by
\beq
\Omega_{\sN_1} h^2 \approx 1.4 \times 10^{-6} \left(\frac{ y_N}{0.5 \times 10^{-13}} \right )^2 \left( \frac{9.5 \textrm{ PeV}}{m_{\tilde h_u}}  \right ) \left( \frac{m_{\sN_1}}{6.5 \textrm{ PeV}} \right ). \label{eq:snrelic}
\eeq
Thus, for PeV masses of the superpartners, the RH sneutrino only makes up a tiny fraction $\mathcal{O}(10^{-6})$ of the energy density of the universe today. We will assume the benchmark value of $y_N = 0.5 \times 10^{-13}$ in the next section.

\section{IceCube neutrino flux}
\label{icecube}
As mentioned in the introduction, the IceCube collaboration has reported 82 ultra-high energy neutrino events with deposited energies in the range from 60 TeV to 10 PeV in their six year data sets~\cite{kopper:2017}. Of these ultra-high energy events, 3 events have been observed with deposited energies greater than 1 PeV. The atmospheric neutrino flux is expected to give a negligible contribution to the event rate above 60 TeV~\cite{Aartsen:2013jdh,Aartsen:2015zva}. Therefore, the ultra-high energy events can be interpreted as strong evidence for either hadronic astrophysical processes ($pp$ or $p\gamma$) or the decay of superheavy DM, or both.

In this paper, we postulate that the PeV scale events observed by the IceCube collaboration arise from the decay of the RH sneutrino, whereas the sub-PeV events arise dominantly from a simple unbroken power law type astrophysical flux. In this section, we calculate the combined neutrino flux from both astrophysical sources and the decay of the RH sneutrino. We will assume the benchmark spectrum for the superpartner masses from the last paragraph of \textsection \ref{sneutrinomassmatrix}. We will also see that in addition to explaining the PeV neutrinos seen at IceCube, the decays of the RH sneutrino could give rise to a flux of very high energy gamma-rays.

\subsection{Neutrino flux from astrophysical sources}
Neutrinos are produced in extreme astrophysical environments involving the interactions of high-energy cosmic rays with photons or other massive particles. Several candidate astrophysical sources such as extragalactic Supernova Remnants (SNRs)~\cite{Murase:2013rfa,Tamborra:2014xia,Chakraborty:2015sta}, Hypernova Remnants~(HNRs)~\cite{Murase:2013rfa,Tamborra:2014xia,Chakraborty:2015sta}, Active Galactic Nuclei (AGNs)~\cite{ Murase:2014foa,Dermer:2014vaa,Kimura:2014jba,Kalashev:2014vya}, and gamma-ray bursts (GRBs)~\cite{Waxman:1997ti,Murase:2013ffa} are expected to give a significant contribution to the neutrino flux. If Fermi shock acceleration was the only mechanism responsible for creating neutrinos~\cite{Gaisser:1990vg}, a power law spectrum $E^{-\gamma}$ with $\gamma \sim 2$ would be expected for the neutrino flux.

.

Different models for a purely astrophysical origin of the high energy neutrino flux have been tested by the IceCube collaboration in \cite{Aartsen:2015knd}, while considering different assumptions about the isotropy of the neutrino flux, as well as its flavor-ratio.
The simplest astrophysical assumptions take an isotropic neutrino flux and a flavor-ratio at Earth for $\nu_e:\nu_\mu:\nu_\tau$ of $1:1:1$, and a simple unbroken power law spectrum (UPL) given by
 \beq
 \label{eq:UPL}
\left.E^2_{\nu} \frac{d \phi}{d E_\nu}\right|_{\rm UPL} = J^{\rm UPL}_0 \left(\frac{E_\nu}{100 {\textrm{ TeV}}}\right)^{-\gamma}.
\eeq
Here, $J^{\rm UPL}_0$ is a normalization of the neutrino flux at $100$ TeV and $\gamma$ is the spectral index.
The IceCube collaboration has found a best fit spectral index $\gamma = 0.5 \pm 0.09$ for neutrinos with energies between $25$ TeV to $2.8$ PeV~\cite{Aartsen:2015zva}. They find that this simple power law flux is consistent with their observations.

However, the observations of three high energy events (above 1 PeV) have generated a lot of interest in the particle physics community. Such high energy events, could arise from a new source of neutrinos, such as from dark matter decay. While the data is still insufficient to discriminate between these alternative hypotheses for the origin of the high energy neutrinos, it is interesting to speculate on a possible dark matter origin that future evidence could either confirm or disprove.

We will assume that the $(10-100)$ TeV events seen at IceCube arise from an UPL type astrophysical flux with parameters $J^{\rm UPL}_0 =1.8 \times 10^{-8}{\textrm{ GeV}}{\textrm{ cm}}^{-2}{\textrm{ s}}^{-1}{\textrm{ sr}}^{-1}$ and $\gamma=0.9$. Next we will try to explain the origin of the PeV scale neutrino events as arising from RH sneutrino decays.

\subsection{Neutrino flux from sneutrino DM decay}
To calculate the expected neutrino flux from the decay of the PeV scale RH sneutrino, we first need to obtain the neutrino spectrum at the source of decay. The neutrino spectrum for the process $\sN_1 \rightarrow \nu_L h^0 \psi_{3/2}$ is determined by both direct neutrino production in sneutrino decay, as well as from the decay of the Higgs particle in the final state. The neutrino spectrum at the source is therefore given by,
\beq
\label{eq:nusource}
\frac{d N}{d E_{\nu}}  = \frac{1}{\Gamma} \frac{d \Gamma}{dE_{\nu}} + \int \frac{1}{\Gamma} \frac{d \Gamma}{dE_{h^0}} \frac{dN_{\nu}(E_{h^0})}{d E_{\nu}} dE_{h^0},
\eeq
where the first term corresponds to the direct neutrino spectrum (with neutrino energies $E_{\nu}$)  and the second term gives the neutrino spectrum from Higgs decay (where the Higgs has an energy $E_{h^0}$). The factor of $\frac{dN_{\nu}(E_{h^0})}{d E_{\nu}}$ gives the neutrino spectrum from Higgs decays and can be taken from the tabulated values in the PPPC4DMID code~\cite{Cirelli:2010xx} which is based on Pythia~\cite{Sjostrand:2007gs}. In order to evaluate the full spectrum, we need to compute the differential decay widths of the sneutrino $\frac{d \Gamma}{dE_{\nu}}$ and $\frac{d \Gamma}{dE_{h^0}}$, as well as the total decay width ${\Gamma}$.

\begin{figure}
\begin{center}
\includegraphics[width = 0.8\textwidth]{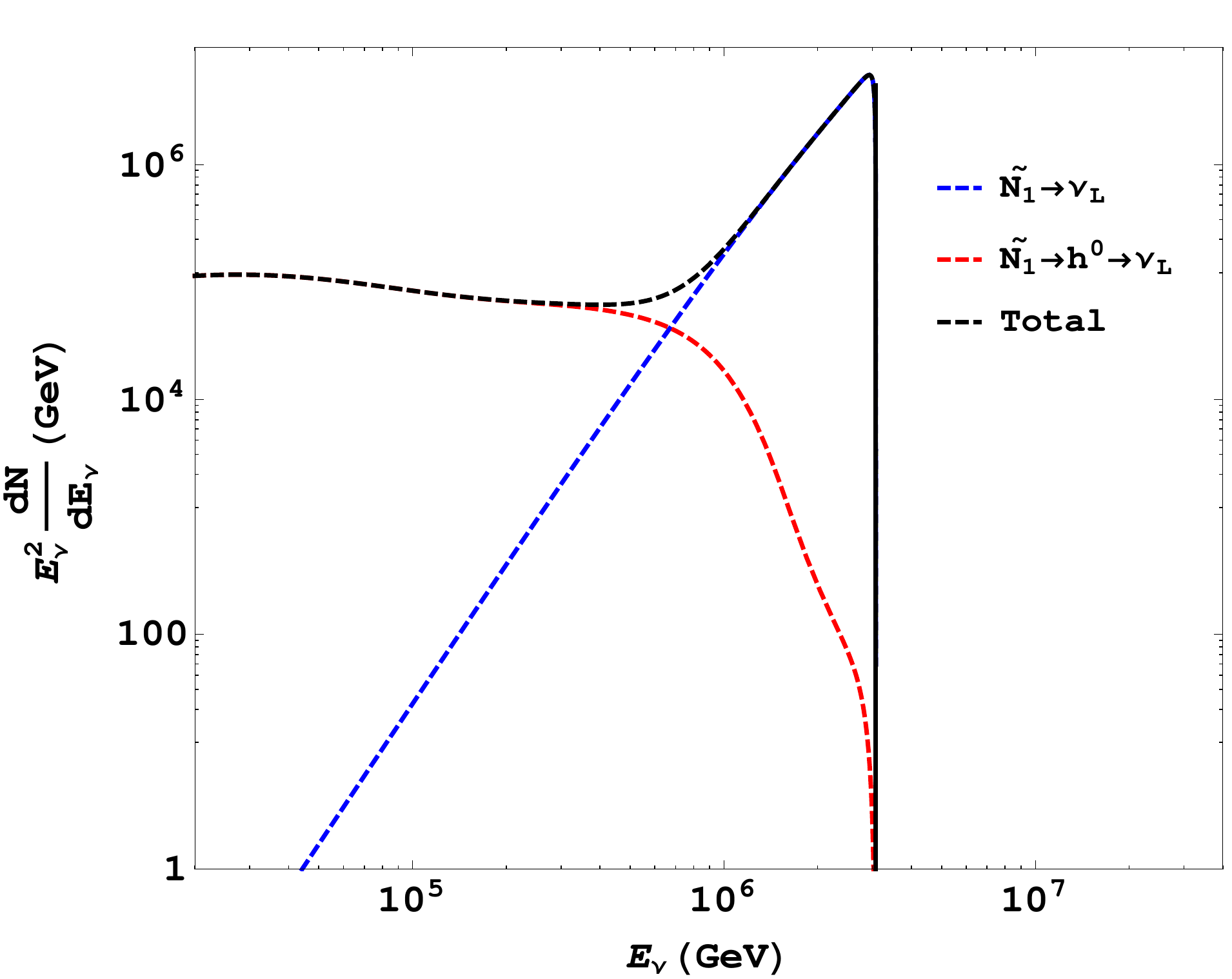}
\caption{Neutrino source spectra from RH sneutrino decay through the process $ \sN_1 \rightarrow  {\psi_{3/2}} \nu_L  h^0$, with $m_{\sN_1} = 6.5$ PeV, $m_N = 8$ PeV and $m_{3/2} = 1.5$ PeV. The dashed blue line shows the neutrino spectrum obtained from the primary neutrinos produced directly from the decay of the RH sneutrino. The red dashed line shows the neutrino spectrum obtained from the decay of the Higgs into neutrinos. Finally, the dashed black line represents the total neutrino spectrum obtained from the three-body decay of the sneutrino.}
\label{nspectra}
\end{center}
\end{figure}

The decay width can be evaluated in terms of the decay amplitude as,
\beq
d \Gamma = \frac{1}{(2 \pi)^3} \frac{1}{8 m_{\sN_1}} \sum_{\rm pol} \left| {\cal M}\right|^2 d E_\nu d E_{h^0}.
\eeq
Here, the spin-summed squared matrix amplitude is given by,
\bea
&& \sum_{\rm pol} \left|{\cal M}\right|^2  \sim \frac{m_{\sN_1}^2 y^2_N}{24 {m^2_{3/2}} {m^4_N} {M^2_p }} \Biggl[  8 {E^3_h} m_{\sN_1} \left(-2 {E_\nu} m_{\sN_1}+m_{\sN_1}^2- m^2_N\right) \nonumber\\
 && +4 {E^2_h} \Biggl\{ -12
   {E^2_\nu} m^2_{\sN_1}+16 {E_\nu} m^3_{\sN_1}-4 {E_\nu} {m^2_{3/2}}m_{\sN_1}-5 m^4_{\sN_1}+3 m^2_{\sN_1}
   {m^2_{3/2}}   \nonumber\\
   && -{m^2_N}\left(4 {E_\nu} m_{\sN_1} -5 m_{\sN_1}^2 +{m^2_{3/2}} \right)\Biggr\} +2 {E_h} \Biggl\{-24
   {E^3 _\nu}m_{\sN_1}^2+52 {E^2_\nu} m_{\sN_1}^3-16 {E^2_\nu} m_{\sN_1} {m^2_{3/2}}\nonumber\\
   && -36 {E_\nu} m_{\sN_1}^4  +28
   {E_\nu} m_{\sN_1}^2 {m^2_{3/2}} + {m^2_N} \left({m^2_{3/2}}(5 m_{\sN_1}-2 {E_\nu})-4 m_{\sN_1} ({E_\nu}-2
   m_{\sN_1}) ({E_\nu}-m_{\sN_1})\right) \nonumber\\
   &&  -2 {E_\nu} {m^4_{3/2}} +8 m_{\sN_1}^5-11 m_{\sN_1}^3 {m^2_{3/2}}+3 m_{\sN_1}
   {m^4_{3/2}}\Biggr\}  \nonumber\\
   && +\left({m^2_{3/2}}-4 ({E_\nu}-m_{\sN_1})^2\right) \left(2 {E_\nu}
   m_{\sN_1}-m_{\sN_1}^2+{m^2_{3/2}}\right)^2 \nonumber\\
   && +{m^2_N} \left(4 m_{\sN_1}^2 (E_\nu-m_{\sN_1})^2+m_{\sN_1} {m^2_{3/2}} (4 {E_\nu}-5
   m_{\sN_1})+{m^4_{3/2}}\right)\Biggr],  \nonumber\\
      \eea
where $E_{3/2}= m_{\sN_1}-E_{h^0}-E_{\nu}$ is the energy of the gravitino. Using this, the differential decay width with respect to $E_{\nu}$ is
\beq
\label{eq:dgamma1}
\frac{d \Gamma}{dE_\nu} =  \frac{1}{(2 \pi)^3} \frac{1}{8 m_{\sN_1}} \int^{E^{\rm max}_{h^0}}_{E^{\rm min}_{h^0}} \sum_{\rm pol}  \left|{\cal M}\right|^2 dE_{h^0},
\eeq
where the limits of integration for fixed $E_\nu$ are
\beq
E^{\rm min}_{h^0}= \left(\frac{m_{\sN_1}}{2} - E_\nu\right) - \frac{m^2_{3/2}}{2 m_{\sN_1}}, \   E^{\rm max}_{h^0}=\frac{m_{\sN_1}}{2} -  \frac{m^2_{3/2}}{2 (m_{\sN_1}-2 E_\nu)}.\nonumber
\eeq
One can now perform the above integration to obtain the differential decay width, however we will not reproduce the full and rather lengthy expression here.

Similarly, the differential decay width with respect to $E_{h^0}$ is given by,
\bea
\label{eq:dgamma2}
&& \frac{d \Gamma}{dE_{h^0}} = \frac{1}{(2 \pi)^3} \frac{1}{8 m_{\sN_1}} \int^{E^{\rm max}_{\nu}}_{E^{\rm min}_{\nu}} \sum_{\rm pol}  \left|{\cal M}\right|^2  dE_{\nu},
\eea
where the limits of integration for fixed $E_{h^0}$ are
\beq
  E^{\rm min}_{\nu}=\left(\frac{m_{\sN_1}}{2} - E_{h^0}\right) - \frac{m^2_{3/2}}{2 m_{\sN_1}} , \ E^{\rm max}_{\nu}=\frac{m_{\sN_1}}{2} -  \frac{m^2_{3/2}}{2 (m_{\sN_1}-2 E_{h^0})}.\nonumber
\eeq

We note that the endpoint for the $E_{\nu}$ or $E_{h^0}$ spectra in the above differential decay widths is given by $\frac{m^2_{\sN_1} - m^2_{3/2}}{2 m_{\sN_1}}$. The total decay width can be computed by integrating either of the above differential decay widths.

We can numerically compute the partial and total decay widths for our benchmark sparticle spectrum ($m_{\sN_1} = 6.5$ PeV, $m_N = 8$ PeV and $m_{3/2} = 1.5$ PeV). The lifetime of $\sN_1$ for our benchmark point is found to be $\tau_{\sN_1} \sim 10^{24}$~sec, consistent with our estimate in Eq.~\ref{eq:snlifetime}. Plugging in the numerical partial decay widths in to Eq.~\ref{eq:nusource}, we obtain the final form of the neutrino source spectrum of the RH sneutrino which is shown in Fig.~\ref{nspectra}.
Using this source spectrum, we will next compute the terrestrial neutrino flux obtained from the decay of sneutrino DM accounting for both galactic and extragalactic contributions.

\paragraph{Extragalactic contribution:} DM decays outside of our galaxy will generate neutrinos at a cosmological distance $\chi(z)$ with energy $E^{\prime}_{\nu}$ and these neutrinos will then be incident on the Earth with redshifted energy $E^{\prime}_\nu/(1+z)$. As we have shown in the previous section, most of the DM relic density is dominated by the gravitino which does not contribute to the neutrino flux. We wish to test the feasibility of the sneutrino as a possible candidate to explain the IceCube PeV-scale neutrino events even though it contributes to only a small fraction of the relic density. The differential neutrino flux observed at Earth from extragalactic RH sneutrino decay is given by \cite{Grefe:2008zz},
\beq
 \frac{d \phi}{d E_\nu}  = \frac{ \Omega_{\sN_1} \rho_c}{4 \pi m_{\sN_1} \tau_{\sN_1}} \int^{\infty}_0 dz \frac{1}{H(z)} \frac{d N_\nu}{d E^\prime_\nu},
\eeq
where $ E^{\prime}_\nu=(1+z) E_{\nu}$ is the energy of the neutrinos at source points which are at a redshift $z$, which gives rise to neutrinos at Earth with an energy $E_\nu$. Here, $m_{\sN_1}$, $\tau_{\sN_1}$ and $\Omega_{\sN_1}$ denote the mass, lifetime and present day energy density fraction respectively of the RH sneutrino and $\rho_c$ is the critical density of the universe. The Hubble parameter $H(z)$ is given by $H(z) = H_0 \sqrt{\Omega_{\Lambda} + \Omega_\textrm{m} (1+z)^3 + \Omega_\textrm{r} (1+z)^4}$ and $\Omega_{\Lambda}$, $\Omega_\textrm{m}$ and $\Omega_\textrm{r}$ correspond to the present day energy fractions of dark energy, matter and radiation respectively.

\begin{figure}
\begin{center}
\includegraphics[width = .9\textwidth]{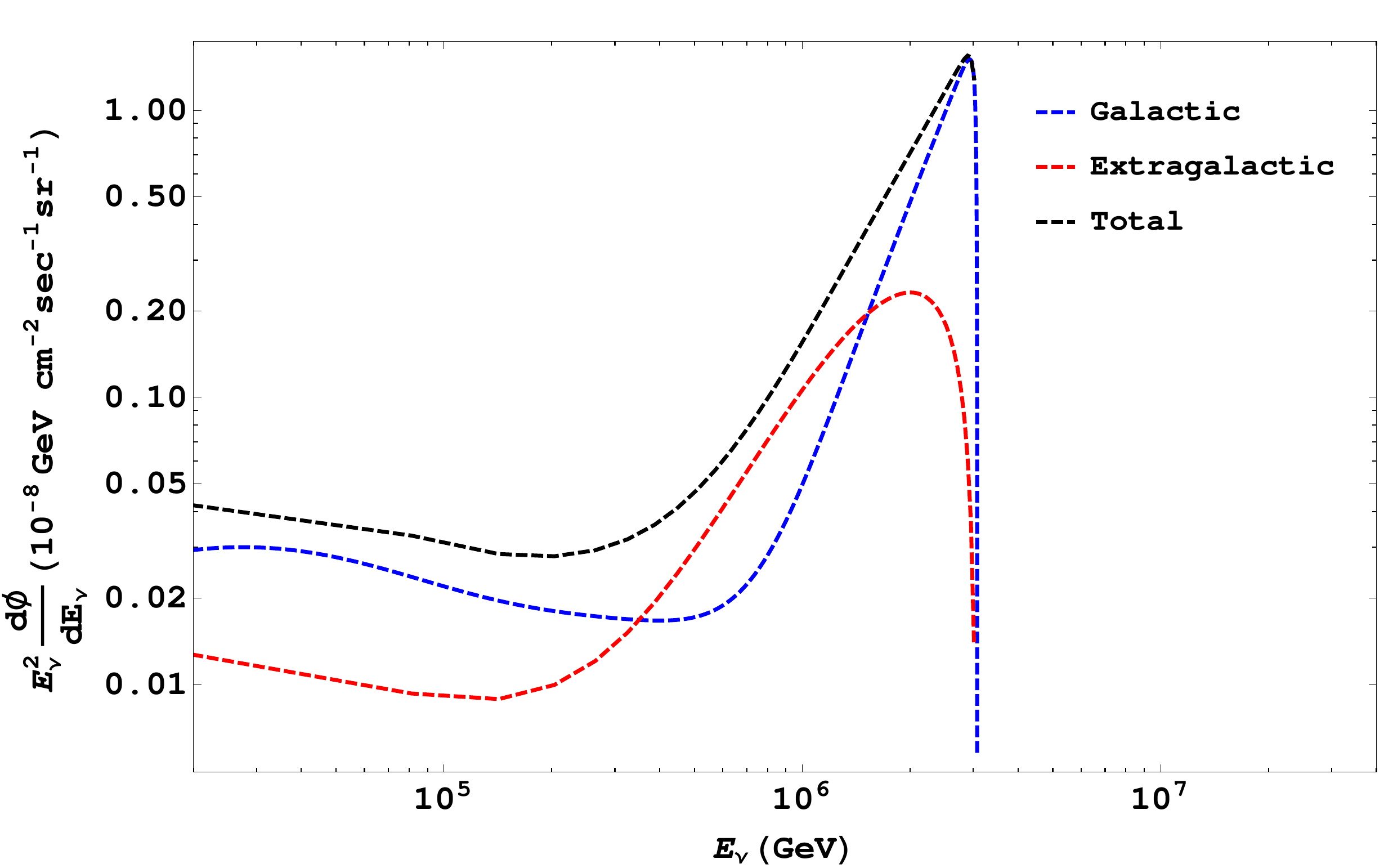}
\caption{ The dashed blue and red lines show the predicted differential neutrino flux at Earth from RH sneutrino decay accounting for galactic and extragalactic contributions, respectively. The dashed black line indicates the total differential neutrino flux which is a sum of these two. We have used $m_{\sN_1} = 6.5$ PeV, $m_N = 8$ PeV, $m_{3/2} = 1.5$ PeV as our benchmark point, which results in  $\tau_{\sN_1} \sim 10^{24}$ sec.}
  \label{fig:extragalactic}
  \end{center}
\end{figure}

\paragraph{Galactic contribution:} Neutrinos from the MilkyWay galactic halo will not experience any cosmological redshift. For the galactic halo, the flux of neutrinos at earth from sneutrino decay per unit energy and time in a volume element located at some point in the halo is given by \cite{Grefe:2008zz},
\beq
 \frac{d \phi}{d E_\nu}  =  \frac{ \Omega_{\sN_1}/\Omega_{\textrm{total DM} } }{4 \pi m_{\rm \sN_1} \tau_{\sN_1}}  \int^{\infty}_0  ds  \ \rho_{\rm DM}(\vec{r})  \ \frac{d N_\nu}{d E_{\nu}},
\eeq

where $s$ is the distance along the line-of-sight from Earth to the DM decay point and $r$ is distance between galactic center and DM decay point which can be written in the form $r = \sqrt{ s^2 + r^{2}_{\odot}- 2 s r_{\odot} \cos{b} \cos{l}}$, where $(s,b,l)$ represent standard galactic coordinates (distance along line-of-sight, latitude and longitude respectively) and $r_{\odot} = 8.5$~kpc is the distance between the earth and the center of the MilkyWay. In the formula above, the factor of $\Omega_{\sN_1}/\Omega_{\textrm{total DM}}$ accounts for the fact that only a fraction of the DM halo density $\rho_{\rm DM}$ is in the form of the RH sneutrino. We assume the density distribution of DM in the galactic halo follows the standard Navarro-Frenk-White (NFW) profile~\cite{Navarro:1995iw}, which is given by
\beq
\rho_{\rm DM}(r) = \frac{\rho_0}{(r/r_c)} \frac{1}{(1+r/r_c)^2},
\eeq
with $r_c = 20$ kpc and $\rho_0 = 0.33$ ${\rm GeV/cm^3}$. Notice that this value of $\rho_0$ accounts for the total DM present in the galactic halo.

\begin{figure}[ht]
\begin{center}
\includegraphics[width = 0.8\textwidth]{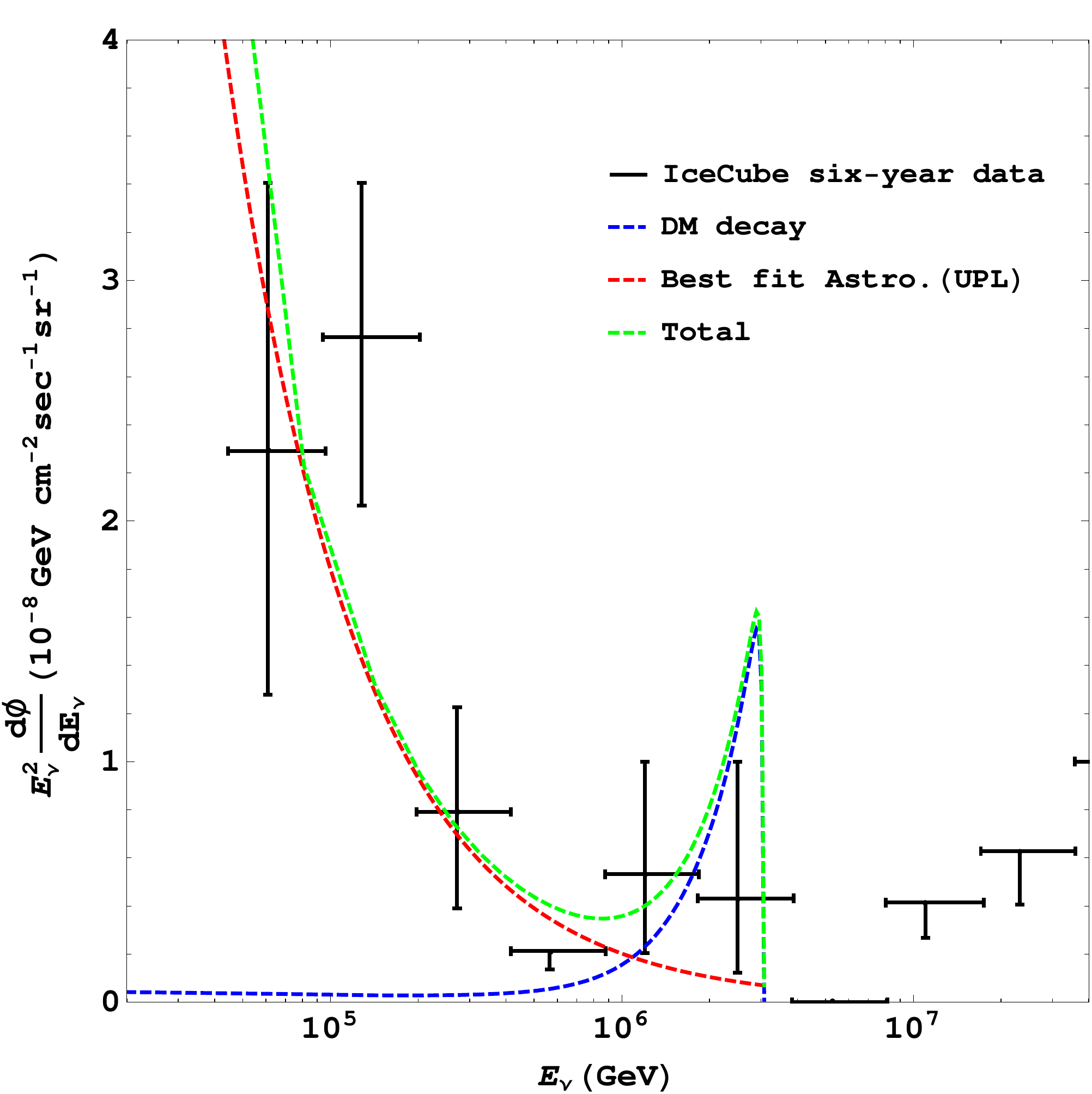}
\end{center}
\caption{This figure shows the total predicted terrestrial neutrino flux coming from both sneutrino decays as well as the nominal UPL astrophysical background. The black data points represent the observed neutrino flux which is inferred from IceCube data~\cite{kopper:2017}. We can see that for our benchmark values $m_{\sN_1} = 6.5$ PeV, $m_N = 8$ PeV and $m_{3/2} = 1.5$ PeV and $\tau_{\sN_1} \sim 10^{24}$ sec, the predicted total neutrino flux closely matches that seen by IceCube, and in particular the PeV events are well accounted for by the sneutrino decays.}
  \label{fig:total}
\end{figure}
\vspace{5mm}

Thus, the total neutrino flux predicted from astrophysics and RH sneutrino decay (including both galactic and extragalactic contributions) is given by,
 \beq
E^2_{\nu} \frac{d \phi}{d E_\nu} =E^2_{\nu} \left[\frac{d \phi_{\rm astro.}}{d E_\nu} + \frac{1}{4 \pi} \int d\Omega \left( \frac{d \phi}{d E_\nu}({\rm galactic}) +  \frac{d \phi}{d E_\nu} ({\rm extragalactic}) \right)\right].
\eeq

For the parameters $\Omega_{\sN_1} =1.5 \times 10^{-6}$, $m_{\sN_1}=6.5$ PeV and $\tau_{\sN_1} \sim 10^{24}$ sec, and taking  values of all the cosmological parameters from the latest results given by Planck~\cite{Ade:2015lrj}, we can compute the final differential flux obtained from sneutrino decay as a sum of galactic and extragalactic contributions. The results are shown in Fig.~\ref{fig:extragalactic}.

In Fig.~\ref{fig:total}, we compare the inferred neutrino flux from six years of IceCube data with our predictions for the total neutrino flux accounting for both astrophysical UPL (given in Eq.~\ref{eq:UPL}) and neutrinos from RH sneutrino decay. From the figure, it is apparent that the sub-PeV flux is best fitted by astrophysical background while the shape of the high energy PeV flux points fit well with the neutrino spectra coming from PeV scale RH sneutrino decay.

Here, we see the second order of magnitude miracle. The lifetime of the RH sneutrino for PeV scale masses of the superpartners is in the range of $10^{24} - 10^{26}$~seconds (see Eq.~\ref{eq:snlifetime}). This combined with the suppressed relic abundance of the RH sneutrino from freeze-in processes (Eq.~\ref{eq:snrelic}), automatically gives us a flux of PeV neutrinos consistent with the flux inferred from IceCube data.

 \begin{figure}
 \begin{center}
 \includegraphics[width = 1.0\textwidth,height=0.8\textwidth]{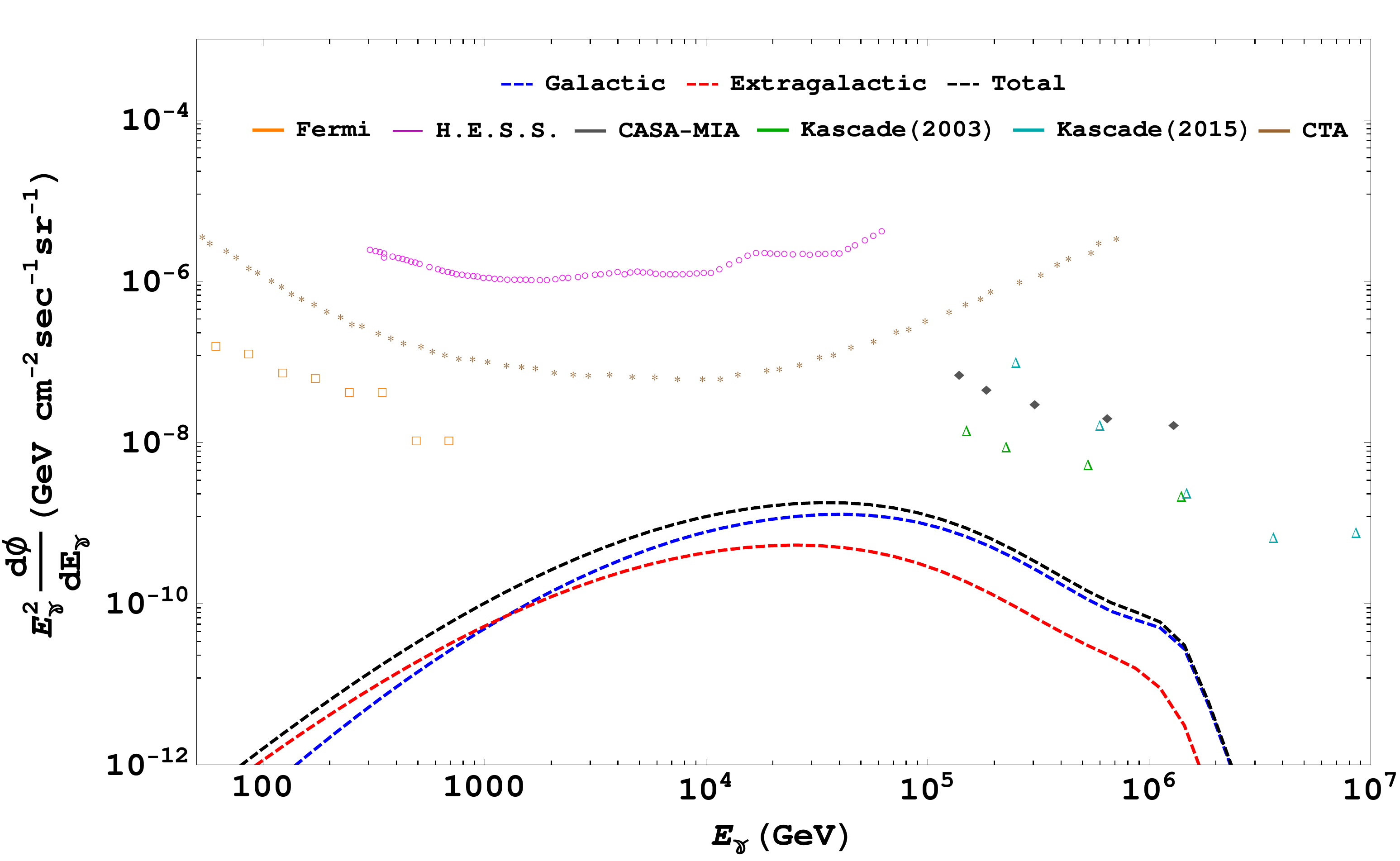}
\caption{The predicted terrestrial gamma-ray flux from the RH sneutrino decay $ \sN_1 \rightarrow  {\psi_{3/2}} \nu_L  h^0$, where the decays of the Higgs give rise to gamma-rays. We have not accounted for secondary gamma-rays here. We have used the benchmark value $m_{\sN_1} = 6.5$~PeV and $m_{3/2} = 1.5$~PeV. The current constraints on the diffuse gamma-ray flux from Fermi-LAT, H.E.S.S., CASA-MIA and KASCADE data are shown, and our predicted flux is well below limits set by these experiments. We have also shown the expected diffuse flux sensitivity of CTA with 500 hours of observation, and we can see that our predicted flux is unfortunately below CTA's sensitivity.}
  \label{fig:gamma}
    \end{center}
\end{figure}

\paragraph{\emph {Additional constraints:}}
Gamma rays signals provide a complementary test of any interpretation of the IceCube data. Previous work has studied the compatibility of gamma ray observations with astrophysical interpretations~\cite{Ahlers:2013xia} as well as decaying DM interpretations of the neutrino flux~\cite{Murase:2015gea,Esmaili:2015xpa,Cohen:2016uyg}. In our model the Higgs produced from RH sneutrino decay can further decay to give prompt as well as secondary high energy gamma-rays. To check the compatibility of expected gamma-ray signals with the observed gamma-ray flux measurements, we calculate the prompt gamma-ray flux spectrum from the decay of sneutrinos by using both galactic and extragalactic contributions. Then we check for consistency of these predictions with gamma-ray bounds from the Fermi-LAT measurement of the isotropic diffuse gamma-ray background~\cite{Ackermann:2014usa}, CASA-MIA~\cite{Borione:1997fy,Prodanovic:2006bq}, the KASCADE experiment~\cite{Schatz:2003aw,Feng:2015dye} and H.E.S.S. measurements~\cite{Rinchiuso:2017kfn}. The comparison between the predicted prompt gamma-ray flux and the constraints from these experiments is shown in Fig.~\ref{fig:gamma}. We can see from the figure that the predicted gamma-ray flux is several orders of magnitude smaller than the observational constraints. We also show in Fig.~\ref{fig:gamma} the predicted diffuse gamma ray flux sensitivity of the Cerenkov Telescope Array (CTA) to a $10^\circ \times 10^\circ$ box around the galactic center with 500 hours of observation. We have made a naive rescaling of the CTA point source sensitivity as given in~\cite{Bernlohr:2012we,Acharya:2017ttl} by assuming background domination in the box, and calibration of the background level by making observations just outside the box. Unfortunately, as we can see from the figure, our predicted gamma-ray flux is below the sensitivity of CTA as well. We note that ref.~\cite{Cohen:2016uyg} showed that the Fermi-LAT bounds for a decaying DM interpretation of the IceCube neutrino flux can be improved by up to one-and-a-half orders of magnitude by using spatial information of the gamma ray flux as well as by using more detailed modelling of the astrophysical sources of gamma rays. However, our low energy gamma ray flux for the benchmark model parameters would still be below this improved sensitivity reach of Fermi.

\section{Summary, conclusions and future prospects}
The search for supersymmetry at the LHC has so far only given us null results. The absence of flavor changing neutral current signals in other experiments have long hinted that the scale of generic SUSY breaking must lie beyond the 100 TeV scale. Moreover the intriguing observation of unexplained PeV scale neutrino events seen at IceCube may be our first direct hint that the SUSY breaking scale is PeV. While this would leave us with a little-hierarchy problem for the Higgs mass, many other attractive features of SUSY such as improved gauge coupling unification would remain.

In this work, we tried to explore the implications of a generic supergravity model with SUSY breaking at the PeV scale. We have built a model with a $U(1)_R$ symmetry with specific $R$-charge assignments, where all the superpartners are at the PeV scale. We have assumed that the gravitino is the LSP and is therefore a good DM candidate. In addition we have assumed that the lightest RH sneutrino $\sN_1$ is the NLSP and in our model, it is also quasi-stable and therefore a candidate for an auxiliary (albeit sub-dominant) DM component. We showed that with our specific $R$-charge assignments, the RH sneutrino has naturally suppressed couplings and the only interactions that it has are Planck-scale suppressed interactions in the K\"{a}hler potential involving the SUSY breaking fields.

We demonstrated that once SUSY, and hence $R$-symmetry is broken, the RH sneutrino is left with suppressed effective interactions to the other particles through trilinear scalar interactions, small mixing with LH sneutrinos and a tiny Yukawa coupling ($y_\nu \sim \textrm{PeV}/M_\textrm{p} \sim 10^{-13}$, see table~\ref{table1}). Of all of these interactions, the Yukawa coupling is the least suppressed and is responsible for both the production of RH sneutrinos in the early universe as well as their eventual decay through the channel $ \sN_1 \rightarrow  {\psi_{3/2}} \nu_L  h^0$, with a very long lifetime much larger than the age of the universe, ${\cal O}(10^{24})$ seconds.

We then calculated the relic abundance of the gravitino and the  RH sneutrino in our model via freeze-in processes.
This is where we encountered our first surprise. \textit{Assuming a reheating tempreature not much larger than the PeV scale, the gravitino abundance automatically had the right order-of-magnitude to make up the bulk of the dark matter of the universe.} The RH sneutrino was found to have a suppressed abundance in comparison, making up only 1 part in $10^{6}$ of the present day density of the universe. This led to the assertion that it makes up a sub-dominant part of the DM density.

The second surprise we encountered was when computing the expected flux of neutrinos observed at Earth from the decay of the RH sneutrino. \textit{We found that even though the RH sneutrino is expected to form a sub-dominant component of the DM relic density, its lifetime is naturally of the right order that it predicts a PeV scale neutrino flux at IceCube consistent with the level that has been observed.} In the literature it has been pointed out that if decaying DM has to explain the IceCube neutrino events, one typically requires the DM to have a lifetime of ${\cal O}(10^{29})-{\cal O}(10^{31})$ seconds to be consistent with the resulting flux inferred from the IceCube data. This is based on the assumption that the decaying DM accounts for the entire DM relic abundance (i.e. $\Omega_{\rm DM}h^2 \sim 0.1$) of the universe while in our case, the lifetime and relic density fraction of the sneutrino turn out to be ${\cal O}(10^{24})$~seconds and $\mathcal{O}(10^{-6})$ respectively. Thus, the suppression in the magnitude of neutrino flux due to suppressed relic abundance is balanced by the larger decay width of the sneutrino, which gives us the correct magnitude of the neutrino flux required to match IceCube observations.

We also showed that in our model, one naturally solves the $\mu$ problem, via the Giudice-Masiero mechanism, and we can also obtain the observed Higgs mass of 125 GeV at the cost of a relative tuning between the weak and PeV scales. Another issue that arose, was whether the observed neutrino masses could arise via the see-saw mechanism, given the suppressed Yukawa coupling of RH sneutrinos. We showed that non-renormalizable K\"{a}hler terms with a suppression by an intermediate scale of $M_* = 10^{10}$~GeV, could explain the observed neutrino masses.

Thus in summary, with the simple assumptions of PeV scale superpartner masses and reheating temperatures not much larger than the PeV scale, we naturally get the right relic abundance for dark matter in the form of gravitinos, and we also naturally predict the right flux for the PeV neutrino events observed at IceCube from RH sneutrino decays.

\textbf{Prospects for verifying this scenario:}
A precise measurement of the neutrino flux spectrum from an improved IceCube detector could potentially resolve the differences between simple astrophysical power law backgrounds and the RH sneutrino decay signal that we have postulated in this work. In addition, if the spatial distribution of the high energy neutrino flux is consistent with a DM origin (such as from clustering of events pointing towards the galactic center or dwarf galaxies), this could bolster the evidence for the RH sneutrino decay as the origin of the PeV neutrino flux seen by IceCube.

Since the three-body decay of the sneutrino gives a Higgs boson in the final state, it is possible that DM will also contribute to the prompt gamma ray flux from the decay of the Higgs into gamma-rays. Therefore we have also computed the possible gamma-ray flux expected from the sneutrino decay. Our results indicate that the prompt gamma-ray flux coming from sneutrino decay is well below the bounds set by high energy gamma-ray observatories. We have performed a naive analysis on the prospect of CTA being able to detect this flux, however it seems unlikely that CTA will be sensitive to the predicted flux of 100 TeV gamma-rays.

PeV scale superpartners would be unobservable at the LHC or any foreseeable high energy collider, but future flavor physics measurements could also be sensitive to generic squark/slepton mixings that are expected in gravity mediated SUSY breaking scenarios.
We do not expect any dark matter direct detection signals, since most of the dark matter is in the form of gravitinos in our scenario and the sub-dominant RH sneutrino has extremely tiny couplings to SM particles.

Finally, we note that due to the suppressed value of the Yukawa couplings, the right-handed neutrino in our model cannot be responsible for vanilla leptogenesis \cite{Fukugita:1986hr}. Therefore one must consider other origins of the matter-antimatter asymmetry in this scenario such as soft leptogenesis \cite{DAmbrosio:2003nfv}.

\section{Acknowledgments}
MD would like to acknowledge Stephan West, Gaurav Tomar and Ujjal Dey for useful clarifications. MD would also like to thank Gaurav Goswami for several useful discussions. VR acknowledges useful discussions with Varun Bhalerao, Ranjan Laha, Pratik Majumdar, Danny Marfatia, Tuhin Roy and Christoph Weniger. Both authors would like to thank the organizers and participants of the ``Blueprints Beyond the Standard
Model'' workshop at TIFR for valuable comments and suggestions. VR would like to thank ICTP, Trieste where a part of this work was completed. VR is supported by a DST-SERB Early Career Research Award (ECR/2017/000040) and an IITB-IRCC seed grant.

\appendix
\section{Supersymmetry breaking mechanism }
\label{appendix1}
In this appendix we briefly discuss a mechanism of supersymmetry and $R$-symmetry breaking. We will see that in addition to the usual $F$-term VEV, this mechanism gives rise to a non-zero VEV for the scalar component of the hidden-sector field. Another interesting feature of the SUSY breaking mechanism presented here is that it also leads to a vanishing cosmological constant.

Typically, supersymmetry is broken in the hidden sector and then communicated to the visible sector via Planck scale suppressed operators (sequestering).  Assuming that the theory has a supersymmetric hidden $SU(N_c)$ gauge theory with $N_f$ pairs of hidden quark superfields $Q$ and ${\bar Q}$ in the fundamental representations $N_c$ and ${\bar N_c}$, respectively, there will be a superpotential interaction of the hidden-sector field with hidden quarks superfields given by \cite{Izawa:1995jg},
\beq
W \supset X Q {\bar Q}.
\eeq
This is consistent with both global R-symmetry as well as axial symmetry. The global axial symmetry and R-symmetry have a hidden QCD anomaly which generates a non-perturbative superpotential for X given by \cite{Izawa:1995jg}:
 \beq
 \label{eq:W1}
W  \supset \lambda \Lambda^2_s X + w,
\eeq
where $\Lambda_s$ corresponds to the scale of strong coupling of the hidden QCD while $w$ is a constant term added to the superpotential. This constant term is necessary in any realistic model to obtain a vanishing cosmological constant after SUSY breaking.

The K\"{a}hler potential for the hidden-sector field is given by:
\beq
 \label{eq:K1}
K \supset X X^{\dagger} -\frac{k}{\Lambda^2_s} \left( X X^\dagger \right)^2.
\eeq
Notice that due to presence of strongly coupled dynamics, the higher dimensional terms in Eq.~\ref{eq:K} are suppressed by $\Lambda_s$, and not by $M_{\textrm{p}}$. Now the effective supergravity potential is given by:
\beq
V = \exp\left(\frac{K}{M^2_{\textrm{p}}}\right) \left( K^{X X^*} \left|\frac{\partial W}{\partial X} + \frac{\partial K}{\partial X} \frac{W}{M^2_{\textrm{p}}}\right|^2 - 3 \frac{\left|W^2\right|}{M^2_{\textrm{p}}}\right).
\eeq
Using Eqs. (\ref{eq:W1}) and (\ref{eq:K1}), the potential takes the form
\bea
&& V = \left(1+  4 k \frac{ X X^*}{\Lambda^2_s} \right) \left| \lambda \Lambda^2_s + X^* \frac{w}{M^2_{\textrm{p}}}\right|^2 - \frac{3}{M^2_{\textrm{p}}}\left|w + \lambda \Lambda^2_s X\right|^2.
\eea
For $X = a + ib$, this reduces to
\bea
V \sim \lambda^2 \Lambda^4_s - \frac{3}{M^2_{\textrm{p}}} w^2 - \frac{4}{M^2_{\textrm{p}}} w \lambda \Lambda^2_s a + 4 k \lambda^2 \Lambda^2_s (a^2 + b^2).
\eea
For $k$ > 0, the minima with vanishing cosmological constant will occur at
\beq
w \approx \frac{1}{\sqrt{3}} \lambda \Lambda^2_s M_{\textrm{p}}, \ \ \ \langle X \rangle \approx \frac{\Lambda^2_s}{2\sqrt{3} k M_{\textrm{p}}},
\label{eq:xvev}
\eeq
and $F_X$ will correspondingly be given by
\beq
F_X= e^{G/2} K^{X \bar X} D_X G \sim \sqrt{3} \frac{w}{M^2_{\textrm{p}}} \equiv \Lambda^2_s,
\eeq
where $G= K + {\rm ln} \left| W\right|^2$. The gravitino mass is given by $m_{3/2} = {F_X}/{M_{\textrm{p}}} \sim \Lambda^2_s/M_{\textrm{p}}$. Plugging this into Eq.~\ref{eq:xvev}, the VEV of the scalar component of the hidden sector superfield is approximately given by $\langle X \rangle \approx m_{3/2}$. We use this value of the VEV while calculating the trilinear soft SUSY breaking parameter for the sneutrino in \textsection \ref{sect:outline}.

\end{document}